\documentclass[12pt]{elsart}

\def\boldsymbol#1{\mbox{\boldmath ${#1}$}}
\def\sub#1{\mbox{\scriptsize\hbox{#1}}}

\begin{document}

\begin{frontmatter}

\title{The role of the pion cloud in electroproduction of 
       the $\Delta$(1232)}

\author[Coimbra,EM1]{M. Fiolhais},
\author[PeF,IJS,EM2]{B. Golli} and
\author[IJS,EM3]{S. \v{S}irca}

\address[Coimbra]{Department of Physics, University of Coimbra,
                  3000 Coimbra, Portugal}
\address[PeF]{Faculty of Education, University of Ljubljana, 
              Ljubljana, Slovenia}
\address[IJS]{J.~Stefan Institute, Jamova 39,
              61111 Ljubljana, Slovenia}

\thanks[EM1]{E-mail: tmanuel@hydra.ci.uc.pt}
\thanks[EM2]{E-mail: bojan.golli@ijs.si}
\thanks[EM3]{E-mail: simon.sirca@ijs.si}

\date{29 January 1996}

\begin{abstract}
We calculate the ratios $E2/M1$ and $C2/M1$ of the multipole
amplitudes for electroproduction of the $\Delta$(1232) in
the range of photon virtuality $0<-K^2<1$~GeV$^2$ in a 
chiral chromodielectric model and a linear $\sigma$-model.
We find that relatively large experimental values can be explained 
in terms of the pion contribution alone; the contribution 
arising from d-state quark admixture remains below 10\%.
We describe the pion cloud as a coherent state and use spin and 
isospin projection to obtain the physical nucleon and the $\Delta$.
The $A_{1/2}$ and $A_{3/2}$ amplitudes are reasonably well
reproduced in the $\sigma$-model; in the chromodielectric  model, 
however, they are a factor of two too small.

\noindent(PACS 12.35H, 13.60P)
\end{abstract}
\end{frontmatter}

The new (e,e$'\pi$) experiments in Mainz and at MIT/Bates,
and those planned at CEBAF have considerably raised the 
interest in theoretical calculations of the amplitudes 
for electroproduction of low lying baryon resonances.
Of particular interest are indications for relatively large 
quadrupole $E2$ (or $E_{1+}$) and $C2$ ($S_{1+}$) amplitudes 
in the vicinity of the $\Delta$(1232) resonance 
which mix with the leading $M1$ ($M_{1+}$) amplitude.
The quark model calculations, assuming d-state quark 
admixtures in the nucleon and the $\Delta$, generally lead 
to much too small values \cite{Karl,Warns,Capstick}.
On the other hand, large values can be reproduced
by assuming a sufficiently strong p-wave pion field 
surrounding three valence quarks.
In the Cloudy Bag Model (CBM), a reasonably good agreement
with the measured amplitudes for photoproduction has been 
obtained with $R=0.6-0.8$~fm \cite{Tiator,Kalb}.
However, at these bag radii the pion field is already so 
strong that the use of perturbative approach is questionable.
It is therefore interesting to study the process in the
framework of a nonperturbative approach such as using
coherent states to describe the pion cloud.

Our aim here is to compute the behaviour of the quadrupole 
amplitudes at low photon virtualities $K^2$ (where the concept 
of the pion as an elementary excitation is still sensible),
and investigate which features do and which do not depend on 
details of a particular model. 
We therefore study the process in the chiral chromodielectric 
model (CDM) and the linear $\sigma$ model (LSM), which, though 
providing different pictures for the nucleon, both account 
for a good description of its static properties \cite{Birse}.
In these models the nucleon and the $\Delta$ are described
as chiral solitons resulting from the non-linear 
interactions between quarks and scalar-isoscalar ($\sigma$)
and pseudoscalar-isovector ($\pol{\pi}$) mesons.
The CDM contains, in addition, a scalar-isoscalar chiral
singlet field $\chi$ which, through the peculiar way it couples 
to the quarks, provides a mechanism for confinement.
In the LSM the pion field in the nucleon is relatively strong 
as a consequence of the topologically nontrivial solution for 
the meson fields resembling in many aspects the Skyrmion 
solution, while in the CDM it is weaker and similar to the 
solution in the CBM (for $R$ above 1~fm).

The Lagrangian of the models can be written as \cite{Birse}
\begin{equation}
  \mathcal{L} = \mathcal{L}_q + \mathcal{L}_{\sigma,\pi} 
    +  \mathcal{L}_{q-\mathrm{meson}} +  \mathcal{L}_\chi\;,
  \label{langrangian}
\end{equation}
where
\begin{equation}
  \mathcal{L}_q = \mathrm{i}\bar{\psi}\gamma^\mu \partial_\mu\psi \;,
\qquad
  \mathcal{L}_{\sigma,\pi} =
  \half\partial_\mu\hat{\sigma}\partial^\mu\hat{\sigma}
  + \half\partial_\mu\hat{\pol{\pi}}\cdot\partial^\mu\hat{\pol{\pi}} 
  - \mathcal{U}(\hat{\pol{\pi}}^2+\hat{\sigma}^2)\;,
  \label{langrangian1}
\end{equation}
$\mathcal{U}(\hat{\pol{\pi}}^2+\hat{\sigma}^2)$ being the usual
Mexican hat potential, and the quark meson interaction is given by
\begin{equation}
    \mathcal{L}_{q-{\mathrm{meson}}} = {g\over\chi^p}\, \bar{\psi}
    (\hat{\sigma}+\mathrm{i}\pol{\tau}\cdot\hat{\pol{\pi}}\gamma_5)
    \psi\;.
   \label{langrangian2}
\end{equation}
In the LSM, $p=0$; in the CDM we take $p=1$.
The last term in (\ref{langrangian}) -- absent in the LSM model --  
contains the kinetic and the potential piece for the $\chi$-field:
\begin{equation}
  \mathcal{L}_\chi =  
  \half\partial_\mu\hat{\chi}\,\partial^\mu\hat{\chi}
  - {1\over2}M_\chi^2\,\hat{\chi}^2\;.
  \label{langrangian3}
\end{equation}
The second term on the RHS is just the mass term for the $\chi$ field.
Other versions of the CDM consider a quartic potential as well as
other powers $p$ in (\ref{langrangian2}).
By taking just the mass term and $p=1$ the confinement is imposed
in the smoothest way, which seems to be the most appropriate 
choice for the quark matter sector of the CDM \cite{Drago}.

Using coherent states to describe the pion and the $\sigma$-meson 
clouds with the ``hedgehog'' ansatz \cite{BB} for the pion 
field and the three valence quarks, the intrinsic state takes 
the form \cite{GR}
\begin{eqnarray}
   |H\rangle &=& \mathcal{N}
   \mathrm{exp}\left\{\sum_m (-1)^{1-m}\int \d k
   \sqrt{2\pi\omega_k/3}\, k\pi(k) a^\dagger_{1m,-m}(k)\right\}
\nonumber\\
   &&\times 
   \mathrm{exp}\left\{\int \d k\sqrt{2\pi\widetilde{\omega}_k}
   \,k\sigma(k) \widetilde{a}^\dagger(k)\right\} 
   \left(b^\dagger_{u\downarrow}-b^\dagger_{d\uparrow}\right)^3
   |0\rangle \;.
\label{hedgehog}
\end{eqnarray}
Here $\widetilde{a}^\dagger(k)$ and $a^\dagger_{lmt}(k)$ are the 
creation operators for the $\sigma$-meson and the pion, respectively, 
in the spherical basis and $t$ is the third component of isospin, 
$\omega_k^2 = k^2 + m_\pi^2$ and 
$\widetilde{\omega}_k^2 = k^2 + m_\sigma^2$,
$b^\dagger_{u\downarrow}$ ($b^\dagger_{d\uparrow}$)
is the creation operator for the `up' (`down') quark in the
1s-state with the third component of spin $-\half$ ($\half$).
Only s-wave $\sigma$-mesons and p-wave pions are coupled to 
the quark core. 
The functions $\sigma(k)$ and $\pi(k)$ are related 
to the expectation values of the field operators as 
$\langle H|\hat{\sigma}(\vec{r})|H\rangle=\sigma(r)$ and
$\langle H|\hat{\pi}_t(\vec{r})|H\rangle=\pi(r)\hat{r}_{-t}$,
where the profiles $\sigma(r)$ and $\pi(r)$ are the Fourier 
transforms of $\sigma(k)$ and $\pi(k)$, respectively. 
The physical states are obtained by performing 
the Peierls-Yoccoz projection
\begin{equation}
   |J=T;M_T,M_J\rangle = 
   \mathcal{N}'(-1)^{J+M_T}P^J_{M_J,-M_T}|H\rangle\;.
\label{projection}
\end{equation}
In the CDM the $\chi$ field is included in the same way as
the $\sigma$ field and is not affected by projection.
The meson and quark profiles are determined selfconsistently
using variation after projection.

The free parameters of the models have been chosen by requiring
that the calculated static properties of the nucleon agree best 
with the experimental values. In the LSM model we use $g=5.0$.
In the considered version of the CDM the results are predominantly 
sensitive to the quantity $G=\sqrt{gM_\chi}$; we take $G=0.2$~GeV
(and $g=0.03$~GeV). We have checked that our results depend very 
weakly on the variations of these parameters.
The models contain three other parameters, the chiral meson masses 
and the pion decay constant, which are fixed to the following 
values: $m_\pi=0.14$~GeV, $m_\sigma=1.2$~GeV, $f_\pi=0.093$~GeV.

It is known that in these models the $\Delta$-N mass splitting is 
too small (typically 160~MeV in the LSM and only 60~MeV in the CDM).
It has been suggested \cite{Kim} that this deficiency can be cured 
through the `t~Hooft interaction which is attractive for the quarks 
in the bare nucleon and absent for the bare $\Delta$.
In our calculation we include this effect as well as the effect of 
residual chromomagnetic interaction by considering different masses 
for the bare nucleon and $\Delta$ in order to reproduce the physical 
$\Delta$-N splitting. The relevant parameter is the difference 
between the bare masses, $\varepsilon_{\Delta{\sub{N}}}$.
While it has only little effect on the pion clouds in the LSM, in 
the CDM it considerably increases the strength of the pion field in 
the $\Delta$, whereas that in the nucleon is slightly decreased.
This can be easily understood: in the variational calculation of the 
physical $\Delta$ the energetically less favourable bare $\Delta$ 
(otherwise the dominant contribution) is being suppressed in favour 
of the configuration with one (or more) pions around the bare nucleon.

The helicity amplitudes $A_{1/2}$ and $A_{3/2}$ for electroproduction 
of the $\Delta$ resonance are defined as
\begin{equation}
   A_\lambda = -{e\over\sqrt{2k_W}}\,\langle\Delta; 
   \half,\lambda\,\vert
   \int\d^3\vec{r}\,\vec{\epsilon}\cdot\widehat{\vec{J}}(\vec{r})
   \,\e^{\mathrm{i}\mbox{\scriptsize\boldmath $kr$}}\,
   \vert\,\mbox{N};\half,\lambda-1\rangle\;,
   \label{defA}
\end{equation}
where $e=\sqrt{4\pi\alpha}$, 
$\vec{\epsilon}=-1/\sqrt{2}\,(1,\mathrm{i},0)$ and the photon 
three-momentum $\vec{k}$ is given by conservation of energy as
\begin{equation}
  \vert\vec{k}\vert^2=\omega^2-K^2\equiv k^2=
  \biggl[ {M_\Delta^2+M_{\sub{N}}^2-K^2\over 2M_\Delta}\biggr]^2-
  M_{\sub{N}}^2\;.
\label{defk}
\end{equation}
We adopt here the convention \cite{Bourdeau} that for virtual 
photons the factor $1/\sqrt{2\omega}$ in the expansion of the 
photon field is replaced by $1/\sqrt{2k_W}$, $k_W$ being the 
value of $k$ at the photon point ($K^2=0$).%
\footnote[1]{Had we kept the factor $1/\sqrt{2\omega}$ in (7), the 
amplitude would have diverged for $K^2=M_{\sub{N}}^2-M_\Delta^2$. 
Since the amplitude for electroproduction of the $\Delta$ is not 
directly observable, one is allowed to introduce such a convention 
(together with a corresponding convention for the decay amplitude) 
in order to avoid this divergence.}
In practical calculation it is convenient to expand the
operator in (\ref{defA}) in terms of electric and magnetic 
multipoles; the relevant quantities are
\begin{equation}
  M^{\sub{M1}}
  =-{{3\over 2}}\int \d^3\vec{r}\,
  \langle\Delta; \half, \half\vert
  \,(\hat{\vec{r}}\times\widehat{\vec{J}})_1\,\vert
  \mbox{N}; \half, -\half\rangle \,j_1(kr)\;,
\label{defM1}
\end{equation}
\begin{equation}
  M^{\sub{E2}}
  = - {\sqrt{10\pi}\over k}\int\d^3\vec{r}\,
  \langle\Delta;\half,\half\vert
  \left[\vec{\nabla}\times j_2(kr)\vec{Y}_{22}^1(\hat{\vec{r}})\right]
  \cdot\widehat{\vec{J}}(\vec{r})
  \vert\mbox{N};\half,-\half\rangle\,.
\label{defE2}
\end{equation}
The $E2$ multipole (\ref{defE2}) can be expressed in terms of 
the charge operator using current conservation as
\begin{eqnarray}
  M^{\sub{E2}}
  &=&  {\sqrt{15\pi}\over 3}\int\d^3\vec{r}\,
  \langle\Delta;\half,\half\vert\,\biggl[
  {\omega\over k}\,\widehat{\rho}(\vec{r})\,
  {\partial\over\partial r}rj_2(kr)
\nonumber\\
  &&- \mathrm{i}k\,\vec{r}\cdot\widehat{\vec{J}}(\vec{r})\,j_2(kr)
  \biggr]
  \,\vert \mbox{N};\half,-\half\rangle\,Y_{21}(\hat{\vec{r}})\,.
\label{defE2s}
\end{eqnarray}
Defining the Coulomb quadrupole 
\begin{equation}
  M^{\sub{C2}}
  = - \sqrt{20\pi}\int\d^3\vec{r}\,
  \langle\Delta; \half, \half\vert
  \,\widehat{\rho}(\vec{r})\,\vert
  \mbox{N}; \half, \half\rangle
  \,Y_{20}(\hat{\vec{r}})j_2(kr)\,,
\label{defC2}
\end{equation}
the $E2/M1$ and $C2/M1$ ratios are given as
\begin{eqnarray}
  {E2\over M1} &=& {1\over 3}{M^{\sub{E2}}
    \over M^{\sub{M1}}}\,,
\label{e2m1}\\
  {C2\over M1} &=& {1\over 2\sqrt{2}}{M^{\sub{C2}}
    \over M^{\sub{M1}}}\,.
\label{c2m1}
\end{eqnarray}
Note that in the limit $k\to0$, ratios (\ref{e2m1}) and (\ref{c2m1}) 
are equal. The current and the charge density operators contain the 
quark and the pion part:
\begin{eqnarray}
  \widehat{\vec{J}}  &=& 
  \bar{\psi}\vec{\gamma}({\textstyle{1\over6}} + \half\tau_3)\psi
  - (\hat{\pol{\pi}}\times\vec{\nabla}\hat{\pol{\pi}})_3\; ,
\label{current}\\
  \widehat{\rho}(\vec{r}) & = &
  \bar{\psi}\gamma_0({\textstyle{1\over6}} + \half\tau_3)\psi
  + (\hat{\pol{\pi}}\times\hat{\pol{P_\pi}})_3\;,
\label{charge}
\end{eqnarray}
where $\hat{\pol{P_\pi}}$ stands for the canonically conjugate field.
Using the grand spin symmetry of the hedgehog, the evaluation
of transition matrix elements between the projected states 
representing the nucleon and the $\Delta$ is considerably simplified; 
details of the calculation technique as well as a discussion of the 
validity of the hedgehog approximation can be found in \cite{CFGR}. 

If we assume, as in (\ref{projection}), that the quarks occupy 
only the lowest s-state, there is no quark contribution to the 
$E2$ and $C2$ amplitudes. However, the interaction 
(\ref{langrangian2}) between the p-wave pions and the quarks 
mixes the s-state and the d-state ($j=3/2$) quarks. It is then 
possible that the $E2$ and $C2$ photons couple directly to quarks 
yielding nonvanishing contributions to the $E2$ and $C2$ amplitudes. 
Such a contribution can be calculated in a straightforward way 
in the CDM; in the LSM this is directly not possible since there 
are no bound d-states in this model. We calculated this effect in 
the CDM by solving the Dirac equation in the background $\chi$ and 
meson fields. Since the interaction responsible for the mixing 
is weak, it is enough to take into account only the lowest excited 
state and calculate the amplitudes in the first order perturbation 
theory. In contrast to the pion contribution, the quark part
is much more sensitive to the detailed structure of the states, 
in particular to the bare $\Delta$-N mass splitting and to the 
energy difference between d and s-states. The latter is typically 
330~MeV which is too small with regard to the energies of low-lying 
nucleon and $\Delta$ excitations. Anyhow, the result represents 
only a small fraction of the pion contribution.

\begin{figure}
\begin{center}
\setlength{\unitlength}{0.240900pt}
\ifx\plotpoint\undefined\newsavebox{\plotpoint}\fi
\begin{picture}(1349,900)(0,0)
\font\gnuplot=cmr10 at 12pt
\gnuplot
\sbox{\plotpoint}{\rule[-0.200pt]{0.400pt}{0.400pt}}%
\put(220.0,495.0){\rule[-0.200pt]{256.558pt}{0.400pt}}
\put(220.0,113.0){\rule[-0.200pt]{4.818pt}{0.400pt}}
\put(198,113){\makebox(0,0)[r]{--0.15}}
\put(1265.0,113.0){\rule[-0.200pt]{4.818pt}{0.400pt}}
\put(220.0,240.0){\rule[-0.200pt]{4.818pt}{0.400pt}}
\put(198,240){\makebox(0,0)[r]{--0.1}}
\put(1265.0,240.0){\rule[-0.200pt]{4.818pt}{0.400pt}}
\put(220.0,368.0){\rule[-0.200pt]{4.818pt}{0.400pt}}
\put(198,368){\makebox(0,0)[r]{--0.05}}
\put(1265.0,368.0){\rule[-0.200pt]{4.818pt}{0.400pt}}
\put(220.0,495.0){\rule[-0.200pt]{4.818pt}{0.400pt}}
\put(198,495){\makebox(0,0)[r]{0}}
\put(1265.0,495.0){\rule[-0.200pt]{4.818pt}{0.400pt}}
\put(220.0,622.0){\rule[-0.200pt]{4.818pt}{0.400pt}}
\put(198,622){\makebox(0,0)[r]{0.05}}
\put(1265.0,622.0){\rule[-0.200pt]{4.818pt}{0.400pt}}
\put(220.0,750.0){\rule[-0.200pt]{4.818pt}{0.400pt}}
\put(198,750){\makebox(0,0)[r]{0.1}}
\put(1265.0,750.0){\rule[-0.200pt]{4.818pt}{0.400pt}}
\put(220.0,877.0){\rule[-0.200pt]{4.818pt}{0.400pt}}
\put(198,877){\makebox(0,0)[r]{0.15}}
\put(1265.0,877.0){\rule[-0.200pt]{4.818pt}{0.400pt}}
\put(268.0,113.0){\rule[-0.200pt]{0.400pt}{4.818pt}}
\put(268,68){\makebox(0,0){0}}
\put(268.0,857.0){\rule[-0.200pt]{0.400pt}{4.818pt}}
\put(462.0,113.0){\rule[-0.200pt]{0.400pt}{4.818pt}}
\put(462,68){\makebox(0,0){0.2}}
\put(462.0,857.0){\rule[-0.200pt]{0.400pt}{4.818pt}}
\put(656.0,113.0){\rule[-0.200pt]{0.400pt}{4.818pt}}
\put(656,68){\makebox(0,0){0.4}}
\put(656.0,857.0){\rule[-0.200pt]{0.400pt}{4.818pt}}
\put(849.0,113.0){\rule[-0.200pt]{0.400pt}{4.818pt}}
\put(849,68){\makebox(0,0){0.6}}
\put(849.0,857.0){\rule[-0.200pt]{0.400pt}{4.818pt}}
\put(1043.0,113.0){\rule[-0.200pt]{0.400pt}{4.818pt}}
\put(1043,68){\makebox(0,0){0.8}}
\put(1043.0,857.0){\rule[-0.200pt]{0.400pt}{4.818pt}}
\put(1237.0,113.0){\rule[-0.200pt]{0.400pt}{4.818pt}}
\put(1237,68){\makebox(0,0){1}}
\put(1237.0,857.0){\rule[-0.200pt]{0.400pt}{4.818pt}}
\put(220.0,113.0){\rule[-0.200pt]{256.558pt}{0.400pt}}
\put(1285.0,113.0){\rule[-0.200pt]{0.400pt}{184.048pt}}
\put(220.0,877.0){\rule[-0.200pt]{256.558pt}{0.400pt}}
\put(45,495){\makebox(0,0){${\displaystyle{E2}\over\displaystyle{M1}}$}}
\put(752,3){\makebox(0,0){$-K^2\,[\mbox{GeV}^2]$}}
\put(220.0,113.0){\rule[-0.200pt]{0.400pt}{184.048pt}}
\put(833,429){\circle*{18}}
\put(1210,638){\circle*{18}}
\put(559,449){\circle*{18}}
\put(559,564){\circle*{18}}
\put(704,523){\circle*{18}}
\put(849,543){\circle*{18}}
\put(1004,569){\circle*{18}}
\put(1237,373){\circle*{18}}
\put(313,266){\circle*{18}}
\put(394,648){\circle*{18}}
\put(508,393){\circle*{18}}
\put(660,202){\circle*{18}}
\put(1256,699){\circle*{18}}
\put(833.0,393.0){\rule[-0.200pt]{0.400pt}{17.104pt}}
\put(823.0,393.0){\rule[-0.200pt]{4.818pt}{0.400pt}}
\put(823.0,464.0){\rule[-0.200pt]{4.818pt}{0.400pt}}
\put(1210.0,582.0){\rule[-0.200pt]{0.400pt}{26.981pt}}
\put(1200.0,582.0){\rule[-0.200pt]{4.818pt}{0.400pt}}
\put(1200.0,694.0){\rule[-0.200pt]{4.818pt}{0.400pt}}
\put(559.0,385.0){\rule[-0.200pt]{0.400pt}{30.835pt}}
\put(549.0,385.0){\rule[-0.200pt]{4.818pt}{0.400pt}}
\put(549.0,513.0){\rule[-0.200pt]{4.818pt}{0.400pt}}
\put(559.0,536.0){\rule[-0.200pt]{0.400pt}{13.490pt}}
\put(549.0,536.0){\rule[-0.200pt]{4.818pt}{0.400pt}}
\put(549.0,592.0){\rule[-0.200pt]{4.818pt}{0.400pt}}
\put(704.0,487.0){\rule[-0.200pt]{0.400pt}{17.345pt}}
\put(694.0,487.0){\rule[-0.200pt]{4.818pt}{0.400pt}}
\put(694.0,559.0){\rule[-0.200pt]{4.818pt}{0.400pt}}
\put(849.0,505.0){\rule[-0.200pt]{0.400pt}{18.549pt}}
\put(839.0,505.0){\rule[-0.200pt]{4.818pt}{0.400pt}}
\put(839.0,582.0){\rule[-0.200pt]{4.818pt}{0.400pt}}
\put(1004.0,528.0){\rule[-0.200pt]{0.400pt}{19.754pt}}
\put(994.0,528.0){\rule[-0.200pt]{4.818pt}{0.400pt}}
\put(994.0,610.0){\rule[-0.200pt]{4.818pt}{0.400pt}}
\put(1237.0,319.0){\rule[-0.200pt]{0.400pt}{25.776pt}}
\put(1227.0,319.0){\rule[-0.200pt]{4.818pt}{0.400pt}}
\put(1227.0,426.0){\rule[-0.200pt]{4.818pt}{0.400pt}}
\put(313.0,197.0){\rule[-0.200pt]{0.400pt}{33.244pt}}
\put(303.0,197.0){\rule[-0.200pt]{4.818pt}{0.400pt}}
\put(303.0,335.0){\rule[-0.200pt]{4.818pt}{0.400pt}}
\put(394.0,571.0){\rule[-0.200pt]{0.400pt}{36.858pt}}
\put(384.0,571.0){\rule[-0.200pt]{4.818pt}{0.400pt}}
\put(384.0,724.0){\rule[-0.200pt]{4.818pt}{0.400pt}}
\put(508.0,342.0){\rule[-0.200pt]{0.400pt}{24.572pt}}
\put(498.0,342.0){\rule[-0.200pt]{4.818pt}{0.400pt}}
\put(498.0,444.0){\rule[-0.200pt]{4.818pt}{0.400pt}}
\put(660.0,113.0){\rule[-0.200pt]{0.400pt}{49.144pt}}
\put(650.0,113.0){\rule[-0.200pt]{4.818pt}{0.400pt}}
\put(650.0,317.0){\rule[-0.200pt]{4.818pt}{0.400pt}}
\put(1256.0,653.0){\rule[-0.200pt]{0.400pt}{22.163pt}}
\put(1246.0,653.0){\rule[-0.200pt]{4.818pt}{0.400pt}}
\put(1246.0,745.0){\rule[-0.200pt]{4.818pt}{0.400pt}}
\put(268,457){\circle*{12}}
\put(268.0,447.0){\rule[-0.200pt]{0.400pt}{4.818pt}}
\put(258.0,447.0){\rule[-0.200pt]{4.818pt}{0.400pt}}
\put(258.0,467.0){\rule[-0.200pt]{4.818pt}{0.400pt}}
\sbox{\plotpoint}{\rule[-0.400pt]{0.800pt}{0.800pt}}%
\put(1155,812){\makebox(0,0)[r]{LSM}}
\put(1177.0,812.0){\rule[-0.400pt]{15.899pt}{0.800pt}}
\put(268,447){\usebox{\plotpoint}}
\put(268,447.34){\rule{4.200pt}{0.800pt}}
\multiput(268.00,445.34)(11.283,4.000){2}{\rule{2.100pt}{0.800pt}}
\put(288,451.34){\rule{4.000pt}{0.800pt}}
\multiput(288.00,449.34)(10.698,4.000){2}{\rule{2.000pt}{0.800pt}}
\put(307,455.34){\rule{4.200pt}{0.800pt}}
\multiput(307.00,453.34)(11.283,4.000){2}{\rule{2.100pt}{0.800pt}}
\put(327,458.84){\rule{4.577pt}{0.800pt}}
\multiput(327.00,457.34)(9.500,3.000){2}{\rule{2.289pt}{0.800pt}}
\put(346,461.84){\rule{4.577pt}{0.800pt}}
\multiput(346.00,460.34)(9.500,3.000){2}{\rule{2.289pt}{0.800pt}}
\put(365,464.84){\rule{4.818pt}{0.800pt}}
\multiput(365.00,463.34)(10.000,3.000){2}{\rule{2.409pt}{0.800pt}}
\put(385,467.84){\rule{4.577pt}{0.800pt}}
\multiput(385.00,466.34)(9.500,3.000){2}{\rule{2.289pt}{0.800pt}}
\put(404,470.34){\rule{4.577pt}{0.800pt}}
\multiput(404.00,469.34)(9.500,2.000){2}{\rule{2.289pt}{0.800pt}}
\put(423,472.84){\rule{4.818pt}{0.800pt}}
\multiput(423.00,471.34)(10.000,3.000){2}{\rule{2.409pt}{0.800pt}}
\put(443,475.34){\rule{4.577pt}{0.800pt}}
\multiput(443.00,474.34)(9.500,2.000){2}{\rule{2.289pt}{0.800pt}}
\put(462,477.34){\rule{4.577pt}{0.800pt}}
\multiput(462.00,476.34)(9.500,2.000){2}{\rule{2.289pt}{0.800pt}}
\put(481,479.34){\rule{4.818pt}{0.800pt}}
\multiput(481.00,478.34)(10.000,2.000){2}{\rule{2.409pt}{0.800pt}}
\put(501,481.34){\rule{4.577pt}{0.800pt}}
\multiput(501.00,480.34)(9.500,2.000){2}{\rule{2.289pt}{0.800pt}}
\put(520,483.34){\rule{4.818pt}{0.800pt}}
\multiput(520.00,482.34)(10.000,2.000){2}{\rule{2.409pt}{0.800pt}}
\put(540,484.84){\rule{4.577pt}{0.800pt}}
\multiput(540.00,484.34)(9.500,1.000){2}{\rule{2.289pt}{0.800pt}}
\put(559,486.34){\rule{4.577pt}{0.800pt}}
\multiput(559.00,485.34)(9.500,2.000){2}{\rule{2.289pt}{0.800pt}}
\put(578,487.84){\rule{4.818pt}{0.800pt}}
\multiput(578.00,487.34)(10.000,1.000){2}{\rule{2.409pt}{0.800pt}}
\put(598,489.34){\rule{4.577pt}{0.800pt}}
\multiput(598.00,488.34)(9.500,2.000){2}{\rule{2.289pt}{0.800pt}}
\put(617,490.84){\rule{4.577pt}{0.800pt}}
\multiput(617.00,490.34)(9.500,1.000){2}{\rule{2.289pt}{0.800pt}}
\put(636,491.84){\rule{4.818pt}{0.800pt}}
\multiput(636.00,491.34)(10.000,1.000){2}{\rule{2.409pt}{0.800pt}}
\put(656,493.34){\rule{4.577pt}{0.800pt}}
\multiput(656.00,492.34)(9.500,2.000){2}{\rule{2.289pt}{0.800pt}}
\put(675,494.84){\rule{4.577pt}{0.800pt}}
\multiput(675.00,494.34)(9.500,1.000){2}{\rule{2.289pt}{0.800pt}}
\put(694,495.84){\rule{4.818pt}{0.800pt}}
\multiput(694.00,495.34)(10.000,1.000){2}{\rule{2.409pt}{0.800pt}}
\put(714,496.84){\rule{4.577pt}{0.800pt}}
\multiput(714.00,496.34)(9.500,1.000){2}{\rule{2.289pt}{0.800pt}}
\put(753,497.84){\rule{4.577pt}{0.800pt}}
\multiput(753.00,497.34)(9.500,1.000){2}{\rule{2.289pt}{0.800pt}}
\put(772,498.84){\rule{4.577pt}{0.800pt}}
\multiput(772.00,498.34)(9.500,1.000){2}{\rule{2.289pt}{0.800pt}}
\put(733.0,499.0){\rule[-0.400pt]{4.818pt}{0.800pt}}
\put(811,499.84){\rule{4.577pt}{0.800pt}}
\multiput(811.00,499.34)(9.500,1.000){2}{\rule{2.289pt}{0.800pt}}
\put(791.0,501.0){\rule[-0.400pt]{4.818pt}{0.800pt}}
\put(849,500.84){\rule{4.818pt}{0.800pt}}
\multiput(849.00,500.34)(10.000,1.000){2}{\rule{2.409pt}{0.800pt}}
\put(830.0,502.0){\rule[-0.400pt]{4.577pt}{0.800pt}}
\put(907,501.84){\rule{4.818pt}{0.800pt}}
\multiput(907.00,501.34)(10.000,1.000){2}{\rule{2.409pt}{0.800pt}}
\put(869.0,503.0){\rule[-0.400pt]{9.154pt}{0.800pt}}
\put(1043,501.84){\rule{4.577pt}{0.800pt}}
\multiput(1043.00,502.34)(9.500,-1.000){2}{\rule{2.289pt}{0.800pt}}
\put(927.0,504.0){\rule[-0.400pt]{27.944pt}{0.800pt}}
\put(1101,500.84){\rule{4.577pt}{0.800pt}}
\multiput(1101.00,501.34)(9.500,-1.000){2}{\rule{2.289pt}{0.800pt}}
\put(1062.0,503.0){\rule[-0.400pt]{9.395pt}{0.800pt}}
\put(1140,499.84){\rule{4.577pt}{0.800pt}}
\multiput(1140.00,500.34)(9.500,-1.000){2}{\rule{2.289pt}{0.800pt}}
\put(1120.0,502.0){\rule[-0.400pt]{4.818pt}{0.800pt}}
\put(1179,498.84){\rule{4.577pt}{0.800pt}}
\multiput(1179.00,499.34)(9.500,-1.000){2}{\rule{2.289pt}{0.800pt}}
\put(1159.0,501.0){\rule[-0.400pt]{4.818pt}{0.800pt}}
\put(1217,497.84){\rule{4.818pt}{0.800pt}}
\multiput(1217.00,498.34)(10.000,-1.000){2}{\rule{2.409pt}{0.800pt}}
\put(1198.0,500.0){\rule[-0.400pt]{4.577pt}{0.800pt}}
\sbox{\plotpoint}{\rule[-0.500pt]{1.000pt}{1.000pt}}%
\put(1155,767){\makebox(0,0)[r]{CDM}}
\multiput(1177,767)(20.756,0.000){4}{\usebox{\plotpoint}}
\put(1243,767){\usebox{\plotpoint}}
\put(268,448){\usebox{\plotpoint}}
\multiput(268,448)(19.880,5.964){2}{\usebox{\plotpoint}}
\multiput(288,454)(20.072,5.282){0}{\usebox{\plotpoint}}
\put(307.96,459.19){\usebox{\plotpoint}}
\put(328.31,463.28){\usebox{\plotpoint}}
\put(348.62,467.55){\usebox{\plotpoint}}
\put(368.94,471.79){\usebox{\plotpoint}}
\put(389.33,475.68){\usebox{\plotpoint}}
\put(409.83,478.92){\usebox{\plotpoint}}
\put(430.34,482.10){\usebox{\plotpoint}}
\put(450.86,485.24){\usebox{\plotpoint}}
\put(471.42,487.99){\usebox{\plotpoint}}
\put(492.07,490.11){\usebox{\plotpoint}}
\put(512.71,492.23){\usebox{\plotpoint}}
\put(533.36,494.34){\usebox{\plotpoint}}
\put(554.01,496.47){\usebox{\plotpoint}}
\put(574.71,497.83){\usebox{\plotpoint}}
\put(595.44,498.87){\usebox{\plotpoint}}
\put(616.17,499.96){\usebox{\plotpoint}}
\multiput(617,500)(20.727,1.091){0}{\usebox{\plotpoint}}
\put(636.90,501.04){\usebox{\plotpoint}}
\put(657.63,502.00){\usebox{\plotpoint}}
\put(678.38,502.18){\usebox{\plotpoint}}
\put(699.11,503.00){\usebox{\plotpoint}}
\put(719.87,503.00){\usebox{\plotpoint}}
\put(740.61,502.62){\usebox{\plotpoint}}
\put(761.35,502.00){\usebox{\plotpoint}}
\put(782.10,501.47){\usebox{\plotpoint}}
\put(802.82,500.41){\usebox{\plotpoint}}
\put(823.55,499.34){\usebox{\plotpoint}}
\put(844.22,497.50){\usebox{\plotpoint}}
\put(864.93,496.20){\usebox{\plotpoint}}
\put(885.48,493.40){\usebox{\plotpoint}}
\put(906.10,491.09){\usebox{\plotpoint}}
\put(926.63,488.06){\usebox{\plotpoint}}
\multiput(927,488)(20.502,-3.237){0}{\usebox{\plotpoint}}
\put(947.13,484.83){\usebox{\plotpoint}}
\put(967.64,481.65){\usebox{\plotpoint}}
\put(987.95,477.38){\usebox{\plotpoint}}
\put(1008.23,472.94){\usebox{\plotpoint}}
\put(1028.29,467.65){\usebox{\plotpoint}}
\put(1048.15,461.64){\usebox{\plotpoint}}
\put(1068.08,455.87){\usebox{\plotpoint}}
\put(1087.53,448.67){\usebox{\plotpoint}}
\put(1106.66,440.62){\usebox{\plotpoint}}
\put(1125.62,432.19){\usebox{\plotpoint}}
\put(1144.14,422.82){\usebox{\plotpoint}}
\put(1162.32,412.84){\usebox{\plotpoint}}
\multiput(1179,402)(16.709,-12.312){2}{\usebox{\plotpoint}}
\put(1211.99,375.48){\usebox{\plotpoint}}
\put(1226.92,361.08){\usebox{\plotpoint}}
\put(1237,351){\usebox{\plotpoint}}
\end{picture}
\end{center}
\caption{The $E2/M1$ ratio calculated in the linear $\sigma$-model
for $g=5.0$ and $\varepsilon_{\Delta{\sub{N}}} = 0.43$~GeV (full line),
and in the chromodielectric model with quartic potential 
for $g=0.03$~GeV, $M_\chi=1.33$~GeV and
$\varepsilon_{\Delta{\sub{N}}} = 0.33$~GeV  (dotted line).
Experimental points are taken from \protect\cite{PDG,Exp1,Exp3}.}
\end{figure}
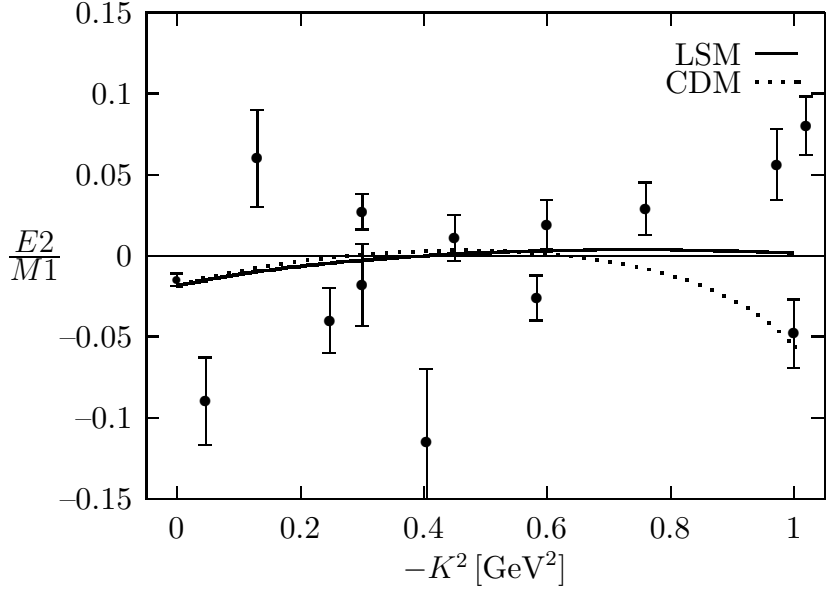
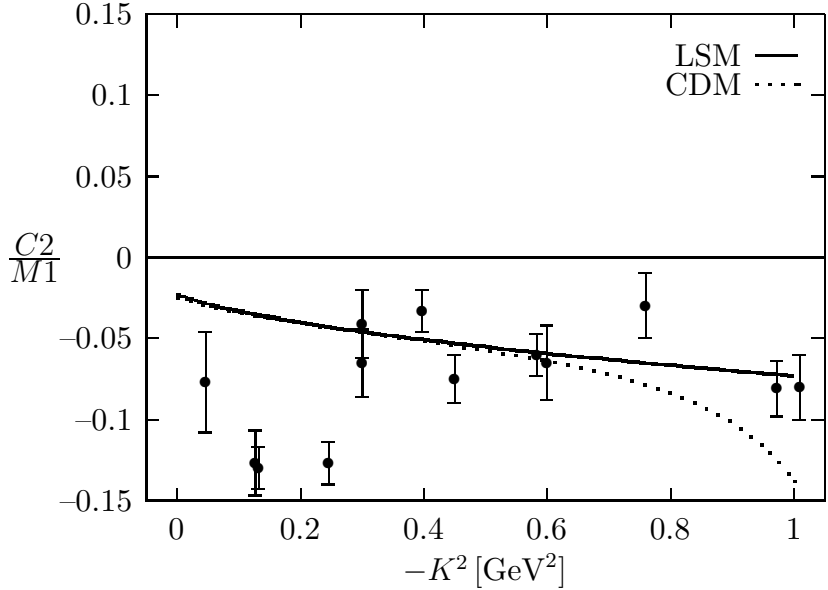
\begin{figure}
\begin{center}
\setlength{\unitlength}{0.240900pt}
\ifx\plotpoint\undefined\newsavebox{\plotpoint}\fi
\sbox{\plotpoint}{\rule[-0.200pt]{0.400pt}{0.400pt}}%
\begin{picture}(1349,900)(0,0)
\font\gnuplot=cmr10 at 12pt
\gnuplot
\sbox{\plotpoint}{\rule[-0.200pt]{0.400pt}{0.400pt}}%
\put(220.0,495.0){\rule[-0.200pt]{256.558pt}{0.400pt}}
\put(220.0,113.0){\rule[-0.200pt]{4.818pt}{0.400pt}}
\put(198,113){\makebox(0,0)[r]{--0.15}}
\put(1265.0,113.0){\rule[-0.200pt]{4.818pt}{0.400pt}}
\put(220.0,240.0){\rule[-0.200pt]{4.818pt}{0.400pt}}
\put(198,240){\makebox(0,0)[r]{--0.1}}
\put(1265.0,240.0){\rule[-0.200pt]{4.818pt}{0.400pt}}
\put(220.0,368.0){\rule[-0.200pt]{4.818pt}{0.400pt}}
\put(198,368){\makebox(0,0)[r]{--0.05}}
\put(1265.0,368.0){\rule[-0.200pt]{4.818pt}{0.400pt}}
\put(220.0,495.0){\rule[-0.200pt]{4.818pt}{0.400pt}}
\put(198,495){\makebox(0,0)[r]{0}}
\put(1265.0,495.0){\rule[-0.200pt]{4.818pt}{0.400pt}}
\put(220.0,622.0){\rule[-0.200pt]{4.818pt}{0.400pt}}
\put(198,622){\makebox(0,0)[r]{0.05}}
\put(1265.0,622.0){\rule[-0.200pt]{4.818pt}{0.400pt}}
\put(220.0,750.0){\rule[-0.200pt]{4.818pt}{0.400pt}}
\put(198,750){\makebox(0,0)[r]{0.1}}
\put(1265.0,750.0){\rule[-0.200pt]{4.818pt}{0.400pt}}
\put(220.0,877.0){\rule[-0.200pt]{4.818pt}{0.400pt}}
\put(198,877){\makebox(0,0)[r]{0.15}}
\put(1265.0,877.0){\rule[-0.200pt]{4.818pt}{0.400pt}}
\put(268.0,113.0){\rule[-0.200pt]{0.400pt}{4.818pt}}
\put(268,68){\makebox(0,0){0}}
\put(268.0,857.0){\rule[-0.200pt]{0.400pt}{4.818pt}}
\put(462.0,113.0){\rule[-0.200pt]{0.400pt}{4.818pt}}
\put(462,68){\makebox(0,0){0.2}}
\put(462.0,857.0){\rule[-0.200pt]{0.400pt}{4.818pt}}
\put(656.0,113.0){\rule[-0.200pt]{0.400pt}{4.818pt}}
\put(656,68){\makebox(0,0){0.4}}
\put(656.0,857.0){\rule[-0.200pt]{0.400pt}{4.818pt}}
\put(849.0,113.0){\rule[-0.200pt]{0.400pt}{4.818pt}}
\put(849,68){\makebox(0,0){0.6}}
\put(849.0,857.0){\rule[-0.200pt]{0.400pt}{4.818pt}}
\put(1043.0,113.0){\rule[-0.200pt]{0.400pt}{4.818pt}}
\put(1043,68){\makebox(0,0){0.8}}
\put(1043.0,857.0){\rule[-0.200pt]{0.400pt}{4.818pt}}
\put(1237.0,113.0){\rule[-0.200pt]{0.400pt}{4.818pt}}
\put(1237,68){\makebox(0,0){1}}
\put(1237.0,857.0){\rule[-0.200pt]{0.400pt}{4.818pt}}
\put(220.0,113.0){\rule[-0.200pt]{256.558pt}{0.400pt}}
\put(1285.0,113.0){\rule[-0.200pt]{0.400pt}{184.048pt}}
\put(220.0,877.0){\rule[-0.200pt]{256.558pt}{0.400pt}}
\put(45,495){\makebox(0,0){${\displaystyle{C2}\over\displaystyle{M1}}$}}
\put(752,3){\makebox(0,0){$-K^2\,[\mbox{GeV}^2]$}}
\put(220.0,113.0){\rule[-0.200pt]{0.400pt}{184.048pt}}
\put(833,342){\circle*{18}}
\put(1210,289){\circle*{18}}
\put(559,391){\circle*{18}}
\put(313,299){\circle*{18}}
\put(396,164){\circle*{18}}
\put(506,172){\circle*{18}}
\put(653,411){\circle*{18}}
\put(559,329){\circle*{18}}
\put(704,304){\circle*{18}}
\put(849,329){\circle*{18}}
\put(1004,419){\circle*{18}}
\put(391,172){\circle*{18}}
\put(1246,291){\circle*{18}}
\put(833.0,309.0){\rule[-0.200pt]{0.400pt}{15.899pt}}
\put(823.0,309.0){\rule[-0.200pt]{4.818pt}{0.400pt}}
\put(823.0,375.0){\rule[-0.200pt]{4.818pt}{0.400pt}}
\put(1210.0,245.0){\rule[-0.200pt]{0.400pt}{20.958pt}}
\put(1200.0,245.0){\rule[-0.200pt]{4.818pt}{0.400pt}}
\put(1200.0,332.0){\rule[-0.200pt]{4.818pt}{0.400pt}}
\put(559.0,337.0){\rule[-0.200pt]{0.400pt}{25.776pt}}
\put(549.0,337.0){\rule[-0.200pt]{4.818pt}{0.400pt}}
\put(549.0,444.0){\rule[-0.200pt]{4.818pt}{0.400pt}}
\put(313.0,220.0){\rule[-0.200pt]{0.400pt}{38.062pt}}
\put(303.0,220.0){\rule[-0.200pt]{4.818pt}{0.400pt}}
\put(303.0,378.0){\rule[-0.200pt]{4.818pt}{0.400pt}}
\put(396.0,131.0){\rule[-0.200pt]{0.400pt}{15.899pt}}
\put(386.0,131.0){\rule[-0.200pt]{4.818pt}{0.400pt}}
\put(386.0,197.0){\rule[-0.200pt]{4.818pt}{0.400pt}}
\put(506.0,138.0){\rule[-0.200pt]{0.400pt}{16.140pt}}
\put(496.0,138.0){\rule[-0.200pt]{4.818pt}{0.400pt}}
\put(496.0,205.0){\rule[-0.200pt]{4.818pt}{0.400pt}}
\put(653.0,378.0){\rule[-0.200pt]{0.400pt}{15.899pt}}
\put(643.0,378.0){\rule[-0.200pt]{4.818pt}{0.400pt}}
\put(643.0,444.0){\rule[-0.200pt]{4.818pt}{0.400pt}}
\put(559.0,276.0){\rule[-0.200pt]{0.400pt}{25.776pt}}
\put(549.0,276.0){\rule[-0.200pt]{4.818pt}{0.400pt}}
\put(549.0,383.0){\rule[-0.200pt]{4.818pt}{0.400pt}}
\put(704.0,266.0){\rule[-0.200pt]{0.400pt}{18.308pt}}
\put(694.0,266.0){\rule[-0.200pt]{4.818pt}{0.400pt}}
\put(694.0,342.0){\rule[-0.200pt]{4.818pt}{0.400pt}}
\put(849.0,271.0){\rule[-0.200pt]{0.400pt}{28.185pt}}
\put(839.0,271.0){\rule[-0.200pt]{4.818pt}{0.400pt}}
\put(839.0,388.0){\rule[-0.200pt]{4.818pt}{0.400pt}}
\put(1004.0,368.0){\rule[-0.200pt]{0.400pt}{24.572pt}}
\put(994.0,368.0){\rule[-0.200pt]{4.818pt}{0.400pt}}
\put(994.0,470.0){\rule[-0.200pt]{4.818pt}{0.400pt}}
\put(391.0,121.0){\rule[-0.200pt]{0.400pt}{24.572pt}}
\put(381.0,121.0){\rule[-0.200pt]{4.818pt}{0.400pt}}
\put(381.0,223.0){\rule[-0.200pt]{4.818pt}{0.400pt}}
\put(1246.0,240.0){\rule[-0.200pt]{0.400pt}{24.572pt}}
\put(1236.0,240.0){\rule[-0.200pt]{4.818pt}{0.400pt}}
\put(1236.0,342.0){\rule[-0.200pt]{4.818pt}{0.400pt}}
\put(268,495){\usebox{\plotpoint}}
\put(268,495){\usebox{\plotpoint}}
\sbox{\plotpoint}{\rule[-0.400pt]{0.800pt}{0.800pt}}%
\put(1155,812){\makebox(0,0)[r]{LSM}}
\put(1177.0,812.0){\rule[-0.400pt]{15.899pt}{0.800pt}}
\put(268,436){\usebox{\plotpoint}}
\multiput(268.00,434.07)(2.025,-0.536){5}{\rule{2.867pt}{0.129pt}}
\multiput(268.00,434.34)(14.050,-6.000){2}{\rule{1.433pt}{0.800pt}}
\multiput(288.00,428.07)(1.913,-0.536){5}{\rule{2.733pt}{0.129pt}}
\multiput(288.00,428.34)(13.327,-6.000){2}{\rule{1.367pt}{0.800pt}}
\multiput(307.00,422.06)(2.943,-0.560){3}{\rule{3.400pt}{0.135pt}}
\multiput(307.00,422.34)(12.943,-5.000){2}{\rule{1.700pt}{0.800pt}}
\put(327,415.34){\rule{4.000pt}{0.800pt}}
\multiput(327.00,417.34)(10.698,-4.000){2}{\rule{2.000pt}{0.800pt}}
\put(346,411.34){\rule{4.000pt}{0.800pt}}
\multiput(346.00,413.34)(10.698,-4.000){2}{\rule{2.000pt}{0.800pt}}
\put(365,407.34){\rule{4.200pt}{0.800pt}}
\multiput(365.00,409.34)(11.283,-4.000){2}{\rule{2.100pt}{0.800pt}}
\put(385,403.34){\rule{4.000pt}{0.800pt}}
\multiput(385.00,405.34)(10.698,-4.000){2}{\rule{2.000pt}{0.800pt}}
\put(404,399.84){\rule{4.577pt}{0.800pt}}
\multiput(404.00,401.34)(9.500,-3.000){2}{\rule{2.289pt}{0.800pt}}
\put(423,396.34){\rule{4.200pt}{0.800pt}}
\multiput(423.00,398.34)(11.283,-4.000){2}{\rule{2.100pt}{0.800pt}}
\put(443,392.84){\rule{4.577pt}{0.800pt}}
\multiput(443.00,394.34)(9.500,-3.000){2}{\rule{2.289pt}{0.800pt}}
\put(462,389.84){\rule{4.577pt}{0.800pt}}
\multiput(462.00,391.34)(9.500,-3.000){2}{\rule{2.289pt}{0.800pt}}
\put(481,386.84){\rule{4.818pt}{0.800pt}}
\multiput(481.00,388.34)(10.000,-3.000){2}{\rule{2.409pt}{0.800pt}}
\put(501,383.84){\rule{4.577pt}{0.800pt}}
\multiput(501.00,385.34)(9.500,-3.000){2}{\rule{2.289pt}{0.800pt}}
\put(520,380.84){\rule{4.818pt}{0.800pt}}
\multiput(520.00,382.34)(10.000,-3.000){2}{\rule{2.409pt}{0.800pt}}
\put(540,378.34){\rule{4.577pt}{0.800pt}}
\multiput(540.00,379.34)(9.500,-2.000){2}{\rule{2.289pt}{0.800pt}}
\put(559,375.84){\rule{4.577pt}{0.800pt}}
\multiput(559.00,377.34)(9.500,-3.000){2}{\rule{2.289pt}{0.800pt}}
\put(578,372.84){\rule{4.818pt}{0.800pt}}
\multiput(578.00,374.34)(10.000,-3.000){2}{\rule{2.409pt}{0.800pt}}
\put(598,370.34){\rule{4.577pt}{0.800pt}}
\multiput(598.00,371.34)(9.500,-2.000){2}{\rule{2.289pt}{0.800pt}}
\put(617,367.84){\rule{4.577pt}{0.800pt}}
\multiput(617.00,369.34)(9.500,-3.000){2}{\rule{2.289pt}{0.800pt}}
\put(636,365.34){\rule{4.818pt}{0.800pt}}
\multiput(636.00,366.34)(10.000,-2.000){2}{\rule{2.409pt}{0.800pt}}
\put(656,363.34){\rule{4.577pt}{0.800pt}}
\multiput(656.00,364.34)(9.500,-2.000){2}{\rule{2.289pt}{0.800pt}}
\put(675,360.84){\rule{4.577pt}{0.800pt}}
\multiput(675.00,362.34)(9.500,-3.000){2}{\rule{2.289pt}{0.800pt}}
\put(694,358.34){\rule{4.818pt}{0.800pt}}
\multiput(694.00,359.34)(10.000,-2.000){2}{\rule{2.409pt}{0.800pt}}
\put(714,356.34){\rule{4.577pt}{0.800pt}}
\multiput(714.00,357.34)(9.500,-2.000){2}{\rule{2.289pt}{0.800pt}}
\put(733,354.34){\rule{4.818pt}{0.800pt}}
\multiput(733.00,355.34)(10.000,-2.000){2}{\rule{2.409pt}{0.800pt}}
\put(753,351.84){\rule{4.577pt}{0.800pt}}
\multiput(753.00,353.34)(9.500,-3.000){2}{\rule{2.289pt}{0.800pt}}
\put(772,349.34){\rule{4.577pt}{0.800pt}}
\multiput(772.00,350.34)(9.500,-2.000){2}{\rule{2.289pt}{0.800pt}}
\put(791,347.34){\rule{4.818pt}{0.800pt}}
\multiput(791.00,348.34)(10.000,-2.000){2}{\rule{2.409pt}{0.800pt}}
\put(811,345.34){\rule{4.577pt}{0.800pt}}
\multiput(811.00,346.34)(9.500,-2.000){2}{\rule{2.289pt}{0.800pt}}
\put(830,343.34){\rule{4.577pt}{0.800pt}}
\multiput(830.00,344.34)(9.500,-2.000){2}{\rule{2.289pt}{0.800pt}}
\put(849,341.34){\rule{4.818pt}{0.800pt}}
\multiput(849.00,342.34)(10.000,-2.000){2}{\rule{2.409pt}{0.800pt}}
\put(869,339.34){\rule{4.577pt}{0.800pt}}
\multiput(869.00,340.34)(9.500,-2.000){2}{\rule{2.289pt}{0.800pt}}
\put(888,337.34){\rule{4.577pt}{0.800pt}}
\multiput(888.00,338.34)(9.500,-2.000){2}{\rule{2.289pt}{0.800pt}}
\put(907,335.84){\rule{4.818pt}{0.800pt}}
\multiput(907.00,336.34)(10.000,-1.000){2}{\rule{2.409pt}{0.800pt}}
\put(927,334.34){\rule{4.577pt}{0.800pt}}
\multiput(927.00,335.34)(9.500,-2.000){2}{\rule{2.289pt}{0.800pt}}
\put(946,332.34){\rule{4.818pt}{0.800pt}}
\multiput(946.00,333.34)(10.000,-2.000){2}{\rule{2.409pt}{0.800pt}}
\put(966,330.34){\rule{4.577pt}{0.800pt}}
\multiput(966.00,331.34)(9.500,-2.000){2}{\rule{2.289pt}{0.800pt}}
\put(985,328.34){\rule{4.577pt}{0.800pt}}
\multiput(985.00,329.34)(9.500,-2.000){2}{\rule{2.289pt}{0.800pt}}
\put(1004,326.34){\rule{4.818pt}{0.800pt}}
\multiput(1004.00,327.34)(10.000,-2.000){2}{\rule{2.409pt}{0.800pt}}
\put(1024,324.84){\rule{4.577pt}{0.800pt}}
\multiput(1024.00,325.34)(9.500,-1.000){2}{\rule{2.289pt}{0.800pt}}
\put(1043,323.34){\rule{4.577pt}{0.800pt}}
\multiput(1043.00,324.34)(9.500,-2.000){2}{\rule{2.289pt}{0.800pt}}
\put(1062,321.34){\rule{4.818pt}{0.800pt}}
\multiput(1062.00,322.34)(10.000,-2.000){2}{\rule{2.409pt}{0.800pt}}
\put(1082,319.84){\rule{4.577pt}{0.800pt}}
\multiput(1082.00,320.34)(9.500,-1.000){2}{\rule{2.289pt}{0.800pt}}
\put(1101,318.34){\rule{4.577pt}{0.800pt}}
\multiput(1101.00,319.34)(9.500,-2.000){2}{\rule{2.289pt}{0.800pt}}
\put(1120,316.34){\rule{4.818pt}{0.800pt}}
\multiput(1120.00,317.34)(10.000,-2.000){2}{\rule{2.409pt}{0.800pt}}
\put(1140,314.34){\rule{4.577pt}{0.800pt}}
\multiput(1140.00,315.34)(9.500,-2.000){2}{\rule{2.289pt}{0.800pt}}
\put(1159,312.84){\rule{4.818pt}{0.800pt}}
\multiput(1159.00,313.34)(10.000,-1.000){2}{\rule{2.409pt}{0.800pt}}
\put(1179,311.34){\rule{4.577pt}{0.800pt}}
\multiput(1179.00,312.34)(9.500,-2.000){2}{\rule{2.289pt}{0.800pt}}
\put(1198,309.84){\rule{4.577pt}{0.800pt}}
\multiput(1198.00,310.34)(9.500,-1.000){2}{\rule{2.289pt}{0.800pt}}
\put(1217,308.34){\rule{4.818pt}{0.800pt}}
\multiput(1217.00,309.34)(10.000,-2.000){2}{\rule{2.409pt}{0.800pt}}
\sbox{\plotpoint}{\rule[-0.500pt]{1.000pt}{1.000pt}}%
\put(1155,767){\makebox(0,0)[r]{CDM}}
\multiput(1177,767)(20.756,0.000){4}{\usebox{\plotpoint}}
\put(1243,767){\usebox{\plotpoint}}
\put(268,432){\usebox{\plotpoint}}
\multiput(268,432)(19.880,-5.964){2}{\usebox{\plotpoint}}
\multiput(288,426)(20.072,-5.282){0}{\usebox{\plotpoint}}
\put(307.95,420.76){\usebox{\plotpoint}}
\put(328.10,415.77){\usebox{\plotpoint}}
\put(348.41,411.49){\usebox{\plotpoint}}
\put(368.73,407.25){\usebox{\plotpoint}}
\put(389.11,403.35){\usebox{\plotpoint}}
\put(409.56,399.83){\usebox{\plotpoint}}
\put(429.94,395.96){\usebox{\plotpoint}}
\put(450.46,392.82){\usebox{\plotpoint}}
\put(470.96,389.59){\usebox{\plotpoint}}
\put(491.47,386.43){\usebox{\plotpoint}}
\put(511.99,383.27){\usebox{\plotpoint}}
\put(532.50,380.12){\usebox{\plotpoint}}
\put(553.10,377.62){\usebox{\plotpoint}}
\put(573.64,374.69){\usebox{\plotpoint}}
\put(594.17,371.58){\usebox{\plotpoint}}
\put(614.67,368.37){\usebox{\plotpoint}}
\put(635.30,366.07){\usebox{\plotpoint}}
\put(655.83,363.03){\usebox{\plotpoint}}
\multiput(656,363)(20.502,-3.237){0}{\usebox{\plotpoint}}
\put(676.33,359.79){\usebox{\plotpoint}}
\put(696.83,356.57){\usebox{\plotpoint}}
\put(717.36,353.47){\usebox{\plotpoint}}
\put(737.86,350.27){\usebox{\plotpoint}}
\put(758.38,347.15){\usebox{\plotpoint}}
\put(778.88,343.91){\usebox{\plotpoint}}
\put(799.40,340.74){\usebox{\plotpoint}}
\put(819.83,337.14){\usebox{\plotpoint}}
\put(840.23,333.38){\usebox{\plotpoint}}
\put(860.65,329.67){\usebox{\plotpoint}}
\put(880.98,325.48){\usebox{\plotpoint}}
\put(901.29,321.20){\usebox{\plotpoint}}
\put(921.47,316.38){\usebox{\plotpoint}}
\put(941.73,311.90){\usebox{\plotpoint}}
\put(961.91,307.02){\usebox{\plotpoint}}
\put(981.77,301.02){\usebox{\plotpoint}}
\put(1001.56,294.77){\usebox{\plotpoint}}
\put(1021.43,288.77){\usebox{\plotpoint}}
\put(1040.96,281.75){\usebox{\plotpoint}}
\put(1060.12,273.79){\usebox{\plotpoint}}
\put(1079.38,266.05){\usebox{\plotpoint}}
\put(1098.21,257.32){\usebox{\plotpoint}}
\put(1116.29,247.15){\usebox{\plotpoint}}
\put(1134.43,237.06){\usebox{\plotpoint}}
\put(1151.59,225.46){\usebox{\plotpoint}}
\put(1168.24,213.07){\usebox{\plotpoint}}
\put(1184.45,200.13){\usebox{\plotpoint}}
\multiput(1198,188)(13.925,-15.391){2}{\usebox{\plotpoint}}
\put(1227.09,154.89){\usebox{\plotpoint}}
\put(1237,143){\usebox{\plotpoint}}
\end{picture}
\end{center}
\caption{The $C2/M1$ ratio; for explanation see Fig.~1.
Experimental points are taken from \protect\cite{Exp1,Exp4}.}
\end{figure}
Our results for the  ratios $E2/M1$ and $C2/M1$ are presented in 
Figs.~1 and 2. The $E2$ amplitude is calculated using (\ref{defE2s}).
For the exact solutions of the model, both (\ref{defE2s}) and 
(\ref{defE2}) should lead to the same result; using approximate 
methods to describe the nucleon and the $\Delta$, it turns out that 
expression (\ref{defE2s}) is much less sensitive to approximations 
than (\ref{defE2}). This finding is well known, particularly in 
nuclear physics as the `Siegert theorem'; in the context of 
electroproduction of the $\Delta$, a similar conclusion has been 
discussed in Ref.~\cite{Bourdeau}. The second term in (\ref{defE2s}) 
is very small and can be neglected; however, substituting
${\partial\over\partial r}rj_2(kr)$ by $3\,j_2(kr)$  as is usually 
done by assuming the small $k$ limit is not justified as a 
consequence of a large contribution from the pion tail.
This is the main origin for different values of $E2/M1$ and $C2/M1$ 
at $K^2=0$ which are otherwise equal in the limit $k\to 0$.
The computed value for $E2/M1$ is $-1.9$\% in the CDM, 
and $-1.8$\% in the LSM, in good agreement with the experimental 
data $-1.5\pm0.4$\% \cite{PDG} -- though this value may be even 
higher in view of the recent measurements in Mainz \cite{Beck}
which give $-2.4\pm0.2$\%. The ratios $C2/M1$ are correspondingly 
$-2.5$\% and $-2.3$\% for the CDM and LSM.
The behaviour of both ratios as a function of $K^2$ is consistent 
with still very uncertain experimental data.
The quark contributions to the quadrupole amplitudes at the photon 
point are less than 10\% with respect to the pion ones,
and raise to around 15\% at $K^2=-1$~GeV$^2$ (for $C2$);
increasing {\it ad hoc\/} the d-s splitting to a more realistic 
value of 500~MeV decreases these values by more than a factor of two.

The situation is not so favourable when considering the absolute 
values of the helicity amplitudes, $A_{1/2}$ and $A_{3/2}$ at $K^2=0$.
In the LSM, the amplitudes (in units of $10^{-3}$~GeV$^{-1/2}$) 
are too small, $-107$ and $-199$, respectively, compared to the 
experimental values $-141\pm 7$ and $-259\pm 10$ \cite{PDG}.
The contribution from pions is 50\%.
In the CDM they are even smaller, $-70$ and $-131$, respectively.
In Fig.~3 we plot the behaviour of 
$M1(K^2)= -[3A_{3/2}+\sqrt{3}A_{1/2}]/2\sqrt{3}$,
which is directly related to the magnetic $\Delta$ transition
form factor $G_{\sub{M}}^*(-K^2)$ \cite{Warns}.
Except for the above discussed discrepancy for $K^2\to 0$,
$M1$ is well reproduced in the LSM, but not in the CDM 
where it rapidly decreases almost to zero at $K^2=-1$~GeV$^2$.
The latter is a consequence of the fact that the $M1$, $E2$ 
and $C2$ transition densities are concentrated at the surface 
(a similar situation occurs also in the MIT model (for $M1$) 
and in the CBM).
At $K^2=-1$~GeV$^2$ the first zeros of $j_1(kr)$ and $j_2(kr)$
already come in the interior of the baryon and cancellation occurs, 
however in such a way that the ratios are almost unaffected.
As our models suggest, the $C2$ amplitude is the most clear 
manifestation of the presence of pions in the baryon; its large 
value at $K^2=-1$~GeV$^2$ may therefore indicate a strong pion 
cloud in the interior.
Such a strong cloud is predicted only in the LSM but not in the CDM.
At $K^2=0$, however, the amplitudes $E2$ and $C2$ only `see' the 
asymptotic pion field whose behaviour is determined by the Yukawa 
form and the $\pi$NN coupling constant, which is reasonably well 
reproduced in both models.
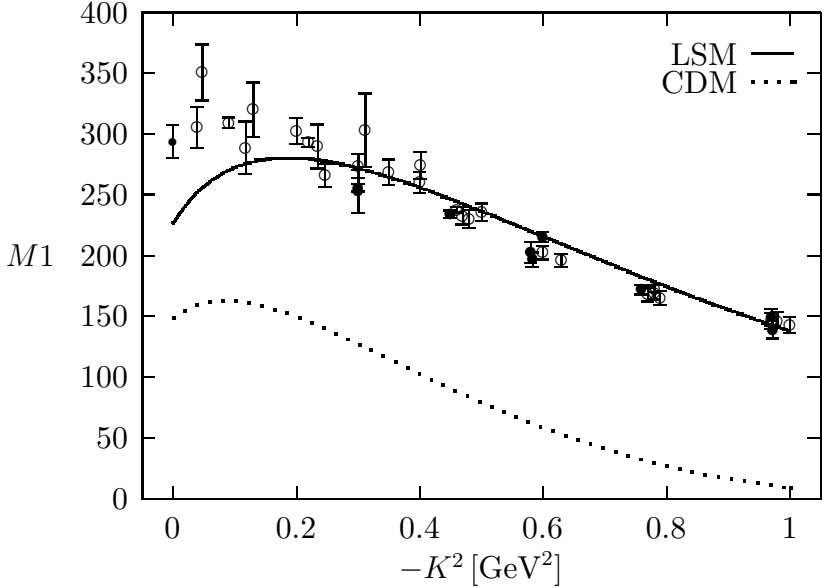
\begin{figure}
\begin{center}
\setlength{\unitlength}{0.240900pt}
\ifx\plotpoint\undefined\newsavebox{\plotpoint}\fi
\sbox{\plotpoint}{\rule[-0.200pt]{0.400pt}{0.400pt}}%
\begin{picture}(1349,900)(0,0)
\font\gnuplot=cmr10 at 12pt
\gnuplot
\sbox{\plotpoint}{\rule[-0.200pt]{0.400pt}{0.400pt}}%
\put(220.0,113.0){\rule[-0.200pt]{256.558pt}{0.400pt}}
\put(220.0,113.0){\rule[-0.200pt]{4.818pt}{0.400pt}}
\put(198,113){\makebox(0,0)[r]{0}}
\put(1265.0,113.0){\rule[-0.200pt]{4.818pt}{0.400pt}}
\put(220.0,209.0){\rule[-0.200pt]{4.818pt}{0.400pt}}
\put(198,209){\makebox(0,0)[r]{50}}
\put(1265.0,209.0){\rule[-0.200pt]{4.818pt}{0.400pt}}
\put(220.0,304.0){\rule[-0.200pt]{4.818pt}{0.400pt}}
\put(198,304){\makebox(0,0)[r]{100}}
\put(1265.0,304.0){\rule[-0.200pt]{4.818pt}{0.400pt}}
\put(220.0,400.0){\rule[-0.200pt]{4.818pt}{0.400pt}}
\put(198,400){\makebox(0,0)[r]{150}}
\put(1265.0,400.0){\rule[-0.200pt]{4.818pt}{0.400pt}}
\put(220.0,495.0){\rule[-0.200pt]{4.818pt}{0.400pt}}
\put(198,495){\makebox(0,0)[r]{200}}
\put(1265.0,495.0){\rule[-0.200pt]{4.818pt}{0.400pt}}
\put(220.0,591.0){\rule[-0.200pt]{4.818pt}{0.400pt}}
\put(198,591){\makebox(0,0)[r]{250}}
\put(1265.0,591.0){\rule[-0.200pt]{4.818pt}{0.400pt}}
\put(220.0,686.0){\rule[-0.200pt]{4.818pt}{0.400pt}}
\put(198,686){\makebox(0,0)[r]{300}}
\put(1265.0,686.0){\rule[-0.200pt]{4.818pt}{0.400pt}}
\put(220.0,782.0){\rule[-0.200pt]{4.818pt}{0.400pt}}
\put(198,782){\makebox(0,0)[r]{350}}
\put(1265.0,782.0){\rule[-0.200pt]{4.818pt}{0.400pt}}
\put(220.0,877.0){\rule[-0.200pt]{4.818pt}{0.400pt}}
\put(198,877){\makebox(0,0)[r]{400}}
\put(1265.0,877.0){\rule[-0.200pt]{4.818pt}{0.400pt}}
\put(268.0,113.0){\rule[-0.200pt]{0.400pt}{4.818pt}}
\put(268,68){\makebox(0,0){0}}
\put(268.0,857.0){\rule[-0.200pt]{0.400pt}{4.818pt}}
\put(462.0,113.0){\rule[-0.200pt]{0.400pt}{4.818pt}}
\put(462,68){\makebox(0,0){0.2}}
\put(462.0,857.0){\rule[-0.200pt]{0.400pt}{4.818pt}}
\put(656.0,113.0){\rule[-0.200pt]{0.400pt}{4.818pt}}
\put(656,68){\makebox(0,0){0.4}}
\put(656.0,857.0){\rule[-0.200pt]{0.400pt}{4.818pt}}
\put(849.0,113.0){\rule[-0.200pt]{0.400pt}{4.818pt}}
\put(849,68){\makebox(0,0){0.6}}
\put(849.0,857.0){\rule[-0.200pt]{0.400pt}{4.818pt}}
\put(1043.0,113.0){\rule[-0.200pt]{0.400pt}{4.818pt}}
\put(1043,68){\makebox(0,0){0.8}}
\put(1043.0,857.0){\rule[-0.200pt]{0.400pt}{4.818pt}}
\put(1237.0,113.0){\rule[-0.200pt]{0.400pt}{4.818pt}}
\put(1237,68){\makebox(0,0){1}}
\put(1237.0,857.0){\rule[-0.200pt]{0.400pt}{4.818pt}}
\put(220.0,113.0){\rule[-0.200pt]{256.558pt}{0.400pt}}
\put(1285.0,113.0){\rule[-0.200pt]{0.400pt}{184.048pt}}
\put(220.0,877.0){\rule[-0.200pt]{256.558pt}{0.400pt}}
\put(45,495){\makebox(0,0){${M1}$}}
\put(752,3){\makebox(0,0){$-K^2\,[\mbox{GeV}^2]$}}
\put(220.0,113.0){\rule[-0.200pt]{0.400pt}{184.048pt}}
\put(268,113){\usebox{\plotpoint}}
\put(268,113){\usebox{\plotpoint}}
\put(268,113){\usebox{\plotpoint}}
\sbox{\plotpoint}{\rule[-0.400pt]{0.800pt}{0.800pt}}%
\put(1155,812){\makebox(0,0)[r]{LSM}}
\put(1177.0,812.0){\rule[-0.400pt]{15.899pt}{0.800pt}}
\put(268,545){\usebox{\plotpoint}}
\multiput(269.41,545.00)(0.505,0.704){33}{\rule{0.122pt}{1.320pt}}
\multiput(266.34,545.00)(20.000,25.260){2}{\rule{0.800pt}{0.660pt}}
\multiput(289.41,573.00)(0.506,0.577){31}{\rule{0.122pt}{1.126pt}}
\multiput(286.34,573.00)(19.000,19.662){2}{\rule{0.800pt}{0.563pt}}
\multiput(307.00,596.41)(0.626,0.507){25}{\rule{1.200pt}{0.122pt}}
\multiput(307.00,593.34)(17.509,16.000){2}{\rule{0.600pt}{0.800pt}}
\multiput(327.00,612.41)(0.740,0.509){19}{\rule{1.369pt}{0.123pt}}
\multiput(327.00,609.34)(16.158,13.000){2}{\rule{0.685pt}{0.800pt}}
\multiput(346.00,625.40)(1.116,0.516){11}{\rule{1.889pt}{0.124pt}}
\multiput(346.00,622.34)(15.080,9.000){2}{\rule{0.944pt}{0.800pt}}
\multiput(365.00,634.40)(1.614,0.526){7}{\rule{2.486pt}{0.127pt}}
\multiput(365.00,631.34)(14.841,7.000){2}{\rule{1.243pt}{0.800pt}}
\multiput(385.00,641.38)(2.775,0.560){3}{\rule{3.240pt}{0.135pt}}
\multiput(385.00,638.34)(12.275,5.000){2}{\rule{1.620pt}{0.800pt}}
\put(404,644.34){\rule{4.577pt}{0.800pt}}
\multiput(404.00,643.34)(9.500,2.000){2}{\rule{2.289pt}{0.800pt}}
\put(423,645.84){\rule{4.818pt}{0.800pt}}
\multiput(423.00,645.34)(10.000,1.000){2}{\rule{2.409pt}{0.800pt}}
\put(462,645.84){\rule{4.577pt}{0.800pt}}
\multiput(462.00,646.34)(9.500,-1.000){2}{\rule{2.289pt}{0.800pt}}
\put(481,643.84){\rule{4.818pt}{0.800pt}}
\multiput(481.00,645.34)(10.000,-3.000){2}{\rule{2.409pt}{0.800pt}}
\put(501,640.84){\rule{4.577pt}{0.800pt}}
\multiput(501.00,642.34)(9.500,-3.000){2}{\rule{2.289pt}{0.800pt}}
\put(520,637.34){\rule{4.200pt}{0.800pt}}
\multiput(520.00,639.34)(11.283,-4.000){2}{\rule{2.100pt}{0.800pt}}
\multiput(540.00,635.06)(2.775,-0.560){3}{\rule{3.240pt}{0.135pt}}
\multiput(540.00,635.34)(12.275,-5.000){2}{\rule{1.620pt}{0.800pt}}
\multiput(559.00,630.06)(2.775,-0.560){3}{\rule{3.240pt}{0.135pt}}
\multiput(559.00,630.34)(12.275,-5.000){2}{\rule{1.620pt}{0.800pt}}
\multiput(578.00,625.07)(2.025,-0.536){5}{\rule{2.867pt}{0.129pt}}
\multiput(578.00,625.34)(14.050,-6.000){2}{\rule{1.433pt}{0.800pt}}
\multiput(598.00,619.07)(1.913,-0.536){5}{\rule{2.733pt}{0.129pt}}
\multiput(598.00,619.34)(13.327,-6.000){2}{\rule{1.367pt}{0.800pt}}
\multiput(617.00,613.07)(1.913,-0.536){5}{\rule{2.733pt}{0.129pt}}
\multiput(617.00,613.34)(13.327,-6.000){2}{\rule{1.367pt}{0.800pt}}
\multiput(636.00,607.08)(1.614,-0.526){7}{\rule{2.486pt}{0.127pt}}
\multiput(636.00,607.34)(14.841,-7.000){2}{\rule{1.243pt}{0.800pt}}
\multiput(656.00,600.08)(1.526,-0.526){7}{\rule{2.371pt}{0.127pt}}
\multiput(656.00,600.34)(14.078,-7.000){2}{\rule{1.186pt}{0.800pt}}
\multiput(675.00,593.08)(1.285,-0.520){9}{\rule{2.100pt}{0.125pt}}
\multiput(675.00,593.34)(14.641,-8.000){2}{\rule{1.050pt}{0.800pt}}
\multiput(694.00,585.08)(1.614,-0.526){7}{\rule{2.486pt}{0.127pt}}
\multiput(694.00,585.34)(14.841,-7.000){2}{\rule{1.243pt}{0.800pt}}
\multiput(714.00,578.08)(1.285,-0.520){9}{\rule{2.100pt}{0.125pt}}
\multiput(714.00,578.34)(14.641,-8.000){2}{\rule{1.050pt}{0.800pt}}
\multiput(733.00,570.08)(1.358,-0.520){9}{\rule{2.200pt}{0.125pt}}
\multiput(733.00,570.34)(15.434,-8.000){2}{\rule{1.100pt}{0.800pt}}
\multiput(753.00,562.08)(1.526,-0.526){7}{\rule{2.371pt}{0.127pt}}
\multiput(753.00,562.34)(14.078,-7.000){2}{\rule{1.186pt}{0.800pt}}
\multiput(772.00,555.08)(1.285,-0.520){9}{\rule{2.100pt}{0.125pt}}
\multiput(772.00,555.34)(14.641,-8.000){2}{\rule{1.050pt}{0.800pt}}
\multiput(791.00,547.08)(1.358,-0.520){9}{\rule{2.200pt}{0.125pt}}
\multiput(791.00,547.34)(15.434,-8.000){2}{\rule{1.100pt}{0.800pt}}
\multiput(811.00,539.08)(1.285,-0.520){9}{\rule{2.100pt}{0.125pt}}
\multiput(811.00,539.34)(14.641,-8.000){2}{\rule{1.050pt}{0.800pt}}
\multiput(830.00,531.08)(1.285,-0.520){9}{\rule{2.100pt}{0.125pt}}
\multiput(830.00,531.34)(14.641,-8.000){2}{\rule{1.050pt}{0.800pt}}
\multiput(849.00,523.08)(1.358,-0.520){9}{\rule{2.200pt}{0.125pt}}
\multiput(849.00,523.34)(15.434,-8.000){2}{\rule{1.100pt}{0.800pt}}
\multiput(869.00,515.08)(1.285,-0.520){9}{\rule{2.100pt}{0.125pt}}
\multiput(869.00,515.34)(14.641,-8.000){2}{\rule{1.050pt}{0.800pt}}
\multiput(888.00,507.08)(1.285,-0.520){9}{\rule{2.100pt}{0.125pt}}
\multiput(888.00,507.34)(14.641,-8.000){2}{\rule{1.050pt}{0.800pt}}
\multiput(907.00,499.08)(1.358,-0.520){9}{\rule{2.200pt}{0.125pt}}
\multiput(907.00,499.34)(15.434,-8.000){2}{\rule{1.100pt}{0.800pt}}
\multiput(927.00,491.08)(1.285,-0.520){9}{\rule{2.100pt}{0.125pt}}
\multiput(927.00,491.34)(14.641,-8.000){2}{\rule{1.050pt}{0.800pt}}
\multiput(946.00,483.08)(1.358,-0.520){9}{\rule{2.200pt}{0.125pt}}
\multiput(946.00,483.34)(15.434,-8.000){2}{\rule{1.100pt}{0.800pt}}
\multiput(966.00,475.08)(1.285,-0.520){9}{\rule{2.100pt}{0.125pt}}
\multiput(966.00,475.34)(14.641,-8.000){2}{\rule{1.050pt}{0.800pt}}
\multiput(985.00,467.08)(1.285,-0.520){9}{\rule{2.100pt}{0.125pt}}
\multiput(985.00,467.34)(14.641,-8.000){2}{\rule{1.050pt}{0.800pt}}
\multiput(1004.00,459.08)(1.614,-0.526){7}{\rule{2.486pt}{0.127pt}}
\multiput(1004.00,459.34)(14.841,-7.000){2}{\rule{1.243pt}{0.800pt}}
\multiput(1024.00,452.08)(1.285,-0.520){9}{\rule{2.100pt}{0.125pt}}
\multiput(1024.00,452.34)(14.641,-8.000){2}{\rule{1.050pt}{0.800pt}}
\multiput(1043.00,444.08)(1.526,-0.526){7}{\rule{2.371pt}{0.127pt}}
\multiput(1043.00,444.34)(14.078,-7.000){2}{\rule{1.186pt}{0.800pt}}
\multiput(1062.00,437.08)(1.358,-0.520){9}{\rule{2.200pt}{0.125pt}}
\multiput(1062.00,437.34)(15.434,-8.000){2}{\rule{1.100pt}{0.800pt}}
\multiput(1082.00,429.08)(1.526,-0.526){7}{\rule{2.371pt}{0.127pt}}
\multiput(1082.00,429.34)(14.078,-7.000){2}{\rule{1.186pt}{0.800pt}}
\multiput(1101.00,422.08)(1.526,-0.526){7}{\rule{2.371pt}{0.127pt}}
\multiput(1101.00,422.34)(14.078,-7.000){2}{\rule{1.186pt}{0.800pt}}
\multiput(1120.00,415.08)(1.614,-0.526){7}{\rule{2.486pt}{0.127pt}}
\multiput(1120.00,415.34)(14.841,-7.000){2}{\rule{1.243pt}{0.800pt}}
\multiput(1140.00,408.08)(1.526,-0.526){7}{\rule{2.371pt}{0.127pt}}
\multiput(1140.00,408.34)(14.078,-7.000){2}{\rule{1.186pt}{0.800pt}}
\multiput(1159.00,401.08)(1.614,-0.526){7}{\rule{2.486pt}{0.127pt}}
\multiput(1159.00,401.34)(14.841,-7.000){2}{\rule{1.243pt}{0.800pt}}
\multiput(1179.00,394.08)(1.526,-0.526){7}{\rule{2.371pt}{0.127pt}}
\multiput(1179.00,394.34)(14.078,-7.000){2}{\rule{1.186pt}{0.800pt}}
\multiput(1198.00,387.07)(1.913,-0.536){5}{\rule{2.733pt}{0.129pt}}
\multiput(1198.00,387.34)(13.327,-6.000){2}{\rule{1.367pt}{0.800pt}}
\multiput(1217.00,381.08)(1.614,-0.526){7}{\rule{2.486pt}{0.127pt}}
\multiput(1217.00,381.34)(14.841,-7.000){2}{\rule{1.243pt}{0.800pt}}
\put(443.0,648.0){\rule[-0.400pt]{4.577pt}{0.800pt}}
\sbox{\plotpoint}{\rule[-0.500pt]{1.000pt}{1.000pt}}%
\put(1155,767){\makebox(0,0)[r]{CDM}}
\multiput(1177,767)(20.756,0.000){4}{\usebox{\plotpoint}}
\put(1243,767){\usebox{\plotpoint}}
\put(268,397){\usebox{\plotpoint}}
\multiput(268,397)(17.402,11.312){2}{\usebox{\plotpoint}}
\put(304.27,416.85){\usebox{\plotpoint}}
\put(324.27,422.32){\usebox{\plotpoint}}
\put(344.91,423.94){\usebox{\plotpoint}}
\multiput(346,424)(20.727,-1.091){0}{\usebox{\plotpoint}}
\put(365.64,422.94){\usebox{\plotpoint}}
\put(386.27,420.73){\usebox{\plotpoint}}
\put(406.51,416.21){\usebox{\plotpoint}}
\put(426.32,410.00){\usebox{\plotpoint}}
\put(446.13,403.85){\usebox{\plotpoint}}
\put(465.55,396.51){\usebox{\plotpoint}}
\put(484.70,388.52){\usebox{\plotpoint}}
\put(503.89,380.63){\usebox{\plotpoint}}
\put(522.68,371.80){\usebox{\plotpoint}}
\put(541.59,363.25){\usebox{\plotpoint}}
\put(560.35,354.36){\usebox{\plotpoint}}
\multiput(578,346)(18.564,-9.282){2}{\usebox{\plotpoint}}
\put(616.41,327.28){\usebox{\plotpoint}}
\put(635.17,318.39){\usebox{\plotpoint}}
\put(653.74,309.13){\usebox{\plotpoint}}
\put(672.47,300.20){\usebox{\plotpoint}}
\put(691.23,291.31){\usebox{\plotpoint}}
\put(710.13,282.74){\usebox{\plotpoint}}
\put(728.93,273.93){\usebox{\plotpoint}}
\put(747.82,265.33){\usebox{\plotpoint}}
\put(766.89,257.15){\usebox{\plotpoint}}
\put(786.02,249.10){\usebox{\plotpoint}}
\put(805.25,241.30){\usebox{\plotpoint}}
\put(824.43,233.35){\usebox{\plotpoint}}
\put(843.80,225.92){\usebox{\plotpoint}}
\put(863.36,218.97){\usebox{\plotpoint}}
\put(882.87,211.89){\usebox{\plotpoint}}
\put(902.34,204.72){\usebox{\plotpoint}}
\put(922.13,198.46){\usebox{\plotpoint}}
\put(941.94,192.28){\usebox{\plotpoint}}
\put(961.80,186.26){\usebox{\plotpoint}}
\put(981.61,180.07){\usebox{\plotpoint}}
\put(1001.64,174.62){\usebox{\plotpoint}}
\put(1021.77,169.56){\usebox{\plotpoint}}
\put(1041.85,164.30){\usebox{\plotpoint}}
\multiput(1043,164)(20.310,-4.276){0}{\usebox{\plotpoint}}
\put(1062.14,159.96){\usebox{\plotpoint}}
\put(1082.28,154.94){\usebox{\plotpoint}}
\put(1102.61,150.75){\usebox{\plotpoint}}
\put(1123.08,147.38){\usebox{\plotpoint}}
\put(1143.46,143.45){\usebox{\plotpoint}}
\put(1163.97,140.25){\usebox{\plotpoint}}
\put(1184.49,137.13){\usebox{\plotpoint}}
\put(1204.99,133.90){\usebox{\plotpoint}}
\put(1225.50,130.72){\usebox{\plotpoint}}
\put(1237,129){\usebox{\plotpoint}}
\sbox{\plotpoint}{\rule[-0.200pt]{0.400pt}{0.400pt}}%
\put(268,674){\circle*{12}}
\put(268.0,649.0){\rule[-0.200pt]{0.400pt}{12.286pt}}
\put(258.0,649.0){\rule[-0.200pt]{4.818pt}{0.400pt}}
\put(258.0,700.0){\rule[-0.200pt]{4.818pt}{0.400pt}}
\put(559,602){\circle*{18}}
\put(704,560){\circle*{18}}
\put(849,525){\circle*{18}}
\put(1004,442){\circle*{18}}
\put(833,489){\circle*{18}}
\put(1210,378){\circle*{18}}
\put(830,500){\circle*{18}}
\put(1208,398){\circle*{18}}
\put(704,560){\circle*{18}}
\put(559,596){\circle*{18}}
\put(559.0,596.0){\rule[-0.200pt]{0.400pt}{2.891pt}}
\put(549.0,596.0){\rule[-0.200pt]{4.818pt}{0.400pt}}
\put(549.0,608.0){\rule[-0.200pt]{4.818pt}{0.400pt}}
\put(704.0,554.0){\rule[-0.200pt]{0.400pt}{2.891pt}}
\put(694.0,554.0){\rule[-0.200pt]{4.818pt}{0.400pt}}
\put(694.0,566.0){\rule[-0.200pt]{4.818pt}{0.400pt}}
\put(849.0,517.0){\rule[-0.200pt]{0.400pt}{3.613pt}}
\put(839.0,517.0){\rule[-0.200pt]{4.818pt}{0.400pt}}
\put(839.0,532.0){\rule[-0.200pt]{4.818pt}{0.400pt}}
\put(1004.0,434.0){\rule[-0.200pt]{0.400pt}{3.613pt}}
\put(994.0,434.0){\rule[-0.200pt]{4.818pt}{0.400pt}}
\put(994.0,449.0){\rule[-0.200pt]{4.818pt}{0.400pt}}
\put(833.0,477.0){\rule[-0.200pt]{0.400pt}{5.541pt}}
\put(823.0,477.0){\rule[-0.200pt]{4.818pt}{0.400pt}}
\put(823.0,500.0){\rule[-0.200pt]{4.818pt}{0.400pt}}
\put(1210.0,365.0){\rule[-0.200pt]{0.400pt}{6.504pt}}
\put(1200.0,365.0){\rule[-0.200pt]{4.818pt}{0.400pt}}
\put(1200.0,392.0){\rule[-0.200pt]{4.818pt}{0.400pt}}
\put(830.0,483.0){\rule[-0.200pt]{0.400pt}{8.191pt}}
\put(820.0,483.0){\rule[-0.200pt]{4.818pt}{0.400pt}}
\put(820.0,517.0){\rule[-0.200pt]{4.818pt}{0.400pt}}
\put(1208.0,384.0){\rule[-0.200pt]{0.400pt}{6.504pt}}
\put(1198.0,384.0){\rule[-0.200pt]{4.818pt}{0.400pt}}
\put(1198.0,411.0){\rule[-0.200pt]{4.818pt}{0.400pt}}
\put(704.0,554.0){\rule[-0.200pt]{0.400pt}{2.891pt}}
\put(694.0,554.0){\rule[-0.200pt]{4.818pt}{0.400pt}}
\put(694.0,566.0){\rule[-0.200pt]{4.818pt}{0.400pt}}
\put(559.0,562.0){\rule[-0.200pt]{0.400pt}{16.381pt}}
\put(549.0,562.0){\rule[-0.200pt]{4.818pt}{0.400pt}}
\put(549.0,630.0){\rule[-0.200pt]{4.818pt}{0.400pt}}
\put(306,697){\circle{18}}
\put(382,664){\circle{18}}
\put(495,667){\circle{18}}
\put(570,692){\circle{18}}
\put(462,691){\circle{18}}
\put(559,636){\circle{18}}
\put(656,610){\circle{18}}
\put(723,558){\circle{18}}
\put(733,553){\circle{18}}
\put(753,563){\circle{18}}
\put(849,500){\circle{18}}
\put(878,488){\circle{18}}
\put(1014,434){\circle{18}}
\put(1024,437){\circle{18}}
\put(1033,428){\circle{18}}
\put(1208,392){\circle{18}}
\put(1217,393){\circle{18}}
\put(607,626){\circle{18}}
\put(1237,386){\circle{18}}
\put(356,704){\circle{18}}
\put(481,673){\circle{18}}
\put(714,567){\circle{18}}
\put(1024,443){\circle{18}}
\put(314,783){\circle{18}}
\put(394,725){\circle{18}}
\put(507,621){\circle{18}}
\put(657,637){\circle{18}}
\put(306.0,664.0){\rule[-0.200pt]{0.400pt}{15.658pt}}
\put(296.0,664.0){\rule[-0.200pt]{4.818pt}{0.400pt}}
\put(296.0,729.0){\rule[-0.200pt]{4.818pt}{0.400pt}}
\put(382.0,623.0){\rule[-0.200pt]{0.400pt}{19.995pt}}
\put(372.0,623.0){\rule[-0.200pt]{4.818pt}{0.400pt}}
\put(372.0,706.0){\rule[-0.200pt]{4.818pt}{0.400pt}}
\put(495.0,632.0){\rule[-0.200pt]{0.400pt}{16.622pt}}
\put(485.0,632.0){\rule[-0.200pt]{4.818pt}{0.400pt}}
\put(485.0,701.0){\rule[-0.200pt]{4.818pt}{0.400pt}}
\put(570.0,634.0){\rule[-0.200pt]{0.400pt}{27.944pt}}
\put(560.0,634.0){\rule[-0.200pt]{4.818pt}{0.400pt}}
\put(560.0,750.0){\rule[-0.200pt]{4.818pt}{0.400pt}}
\put(462.0,671.0){\rule[-0.200pt]{0.400pt}{9.636pt}}
\put(452.0,671.0){\rule[-0.200pt]{4.818pt}{0.400pt}}
\put(452.0,711.0){\rule[-0.200pt]{4.818pt}{0.400pt}}
\put(559.0,617.0){\rule[-0.200pt]{0.400pt}{9.154pt}}
\put(549.0,617.0){\rule[-0.200pt]{4.818pt}{0.400pt}}
\put(549.0,655.0){\rule[-0.200pt]{4.818pt}{0.400pt}}
\put(656.0,593.0){\rule[-0.200pt]{0.400pt}{8.191pt}}
\put(646.0,593.0){\rule[-0.200pt]{4.818pt}{0.400pt}}
\put(646.0,627.0){\rule[-0.200pt]{4.818pt}{0.400pt}}
\put(723.0,544.0){\rule[-0.200pt]{0.400pt}{6.504pt}}
\put(713.0,544.0){\rule[-0.200pt]{4.818pt}{0.400pt}}
\put(713.0,571.0){\rule[-0.200pt]{4.818pt}{0.400pt}}
\put(733.0,539.0){\rule[-0.200pt]{0.400pt}{6.745pt}}
\put(723.0,539.0){\rule[-0.200pt]{4.818pt}{0.400pt}}
\put(723.0,567.0){\rule[-0.200pt]{4.818pt}{0.400pt}}
\put(753.0,549.0){\rule[-0.200pt]{0.400pt}{6.745pt}}
\put(743.0,549.0){\rule[-0.200pt]{4.818pt}{0.400pt}}
\put(743.0,577.0){\rule[-0.200pt]{4.818pt}{0.400pt}}
\put(849.0,489.0){\rule[-0.200pt]{0.400pt}{5.059pt}}
\put(839.0,489.0){\rule[-0.200pt]{4.818pt}{0.400pt}}
\put(839.0,510.0){\rule[-0.200pt]{4.818pt}{0.400pt}}
\put(878.0,477.0){\rule[-0.200pt]{0.400pt}{5.059pt}}
\put(868.0,477.0){\rule[-0.200pt]{4.818pt}{0.400pt}}
\put(868.0,498.0){\rule[-0.200pt]{4.818pt}{0.400pt}}
\put(1014.0,423.0){\rule[-0.200pt]{0.400pt}{5.300pt}}
\put(1004.0,423.0){\rule[-0.200pt]{4.818pt}{0.400pt}}
\put(1004.0,445.0){\rule[-0.200pt]{4.818pt}{0.400pt}}
\put(1024.0,426.0){\rule[-0.200pt]{0.400pt}{5.541pt}}
\put(1014.0,426.0){\rule[-0.200pt]{4.818pt}{0.400pt}}
\put(1014.0,449.0){\rule[-0.200pt]{4.818pt}{0.400pt}}
\put(1033.0,417.0){\rule[-0.200pt]{0.400pt}{5.541pt}}
\put(1023.0,417.0){\rule[-0.200pt]{4.818pt}{0.400pt}}
\put(1023.0,440.0){\rule[-0.200pt]{4.818pt}{0.400pt}}
\put(1208.0,380.0){\rule[-0.200pt]{0.400pt}{6.022pt}}
\put(1198.0,380.0){\rule[-0.200pt]{4.818pt}{0.400pt}}
\put(1198.0,405.0){\rule[-0.200pt]{4.818pt}{0.400pt}}
\put(1217.0,381.0){\rule[-0.200pt]{0.400pt}{6.022pt}}
\put(1207.0,381.0){\rule[-0.200pt]{4.818pt}{0.400pt}}
\put(1207.0,406.0){\rule[-0.200pt]{4.818pt}{0.400pt}}
\put(607.0,606.0){\rule[-0.200pt]{0.400pt}{9.636pt}}
\put(597.0,606.0){\rule[-0.200pt]{4.818pt}{0.400pt}}
\put(597.0,646.0){\rule[-0.200pt]{4.818pt}{0.400pt}}
\put(1237.0,374.0){\rule[-0.200pt]{0.400pt}{6.022pt}}
\put(1227.0,374.0){\rule[-0.200pt]{4.818pt}{0.400pt}}
\put(1227.0,399.0){\rule[-0.200pt]{4.818pt}{0.400pt}}
\put(356.0,696.0){\rule[-0.200pt]{0.400pt}{3.854pt}}
\put(346.0,696.0){\rule[-0.200pt]{4.818pt}{0.400pt}}
\put(346.0,712.0){\rule[-0.200pt]{4.818pt}{0.400pt}}
\put(481.0,666.0){\rule[-0.200pt]{0.400pt}{3.373pt}}
\put(471.0,666.0){\rule[-0.200pt]{4.818pt}{0.400pt}}
\put(471.0,680.0){\rule[-0.200pt]{4.818pt}{0.400pt}}
\put(714.0,563.0){\rule[-0.200pt]{0.400pt}{2.168pt}}
\put(704.0,563.0){\rule[-0.200pt]{4.818pt}{0.400pt}}
\put(704.0,572.0){\rule[-0.200pt]{4.818pt}{0.400pt}}
\put(1024.0,438.0){\rule[-0.200pt]{0.400pt}{2.650pt}}
\put(1014.0,438.0){\rule[-0.200pt]{4.818pt}{0.400pt}}
\put(1014.0,449.0){\rule[-0.200pt]{4.818pt}{0.400pt}}
\put(314.0,739.0){\rule[-0.200pt]{0.400pt}{21.199pt}}
\put(304.0,739.0){\rule[-0.200pt]{4.818pt}{0.400pt}}
\put(304.0,827.0){\rule[-0.200pt]{4.818pt}{0.400pt}}
\put(394.0,682.0){\rule[-0.200pt]{0.400pt}{20.476pt}}
\put(384.0,682.0){\rule[-0.200pt]{4.818pt}{0.400pt}}
\put(384.0,767.0){\rule[-0.200pt]{4.818pt}{0.400pt}}
\put(507.0,603.0){\rule[-0.200pt]{0.400pt}{8.672pt}}
\put(497.0,603.0){\rule[-0.200pt]{4.818pt}{0.400pt}}
\put(497.0,639.0){\rule[-0.200pt]{4.818pt}{0.400pt}}
\put(657.0,616.0){\rule[-0.200pt]{0.400pt}{10.118pt}}
\put(647.0,616.0){\rule[-0.200pt]{4.818pt}{0.400pt}}
\put(647.0,658.0){\rule[-0.200pt]{4.818pt}{0.400pt}}
\end{picture}
\end{center}
\caption{The electromagnetic amplitude 
$M1=-[3A\protect_{3/2}+\protect\sqrt{3}A\protect_{1/2}]/2
\protect\sqrt{3}$
in units of $10\protect^{-3}$~GeV$\protect^{-1/2}$
calculated in the linear $\sigma$-model and
chromodielectric model (see Fig.~1).
The experimental value at the photon point is
taken from \protect\cite{PDG}. 
For virtual photons, the values denoted by $\bullet$ are 
determined from the $|M\protect_{1+}|$ amplitude for the process 
$\gamma\mathrm{p}\to \mathrm{p}\pi^0$ \protect\cite{Exp1} 
assuming the dominance of the $T=3/2$ amplitude,
the values denoted by $\circ$ are computed from 
$G_{\protect\sub{M}}^*$ \protect\cite{Exp3,Exp2}.}
\end{figure}

To the best of our knowledge this is the first calculation of the 
electroproduction quadrupole amplitudes in models with the pion 
cloud; other calculations all refer only to the photoproduction case. 
Apart from the already mentioned calculations in the CBM 
\cite{Tiator,Kalb}, the ratio $E2/M1$ was calculated in the
Skyrme model \cite{Weise} where almost a factor of two larger value 
was obtained. In Ref.~\cite{Christov}, where the NJL model was used,
the effect of the pion cloud does not appear explicitly, but rather 
through the excitations of the Dirac sea.
The value for $E2/M1$ they obtained is very close to our predictions 
(taking into account that they actually used the formula for $C2/M1$).
The constituent quark model has been used to describe both 
photoproduction and electroproduction.
The predicted values for the ratios $E2/M1$ and $C2/M1$ are much 
smaller than the experimental ones -- a clear indication that they 
cannot be reproduced in terms of valence quarks alone.
It is interesting to notice that a good estimate for the $E2/M1$ 
ratio can be obtained in the old CGLN model \cite{CGLN} in which 
the nucleon and the $\Delta$ are described as a bare nucleon 
surrounded by a pion cloud having a Yukawa form with a cut-off in 
momentum space. Using their formulas for $E_{1+}$ and $M_{1+}$ and 
neglecting the crossed term one obtains a value around $-2$\%,
almost insensitive to the cut-off momentum.
This strongly supports our conclusion that in photoproduction
the ratio is essentially determined by the tail of the pion cloud.

It is instructive to analyze the results for the production 
amplitudes in terms of the relation $\mu_{\Delta{\sub{N}}} = 
{2\over3}\mu_{\sub{v}} g_{\pi\Delta{\sub{N}}}/g_{\pi{\sub{NN}}}$,
where the $g$'s are the $\pi\Delta$N and $\pi$NN coupling constants, 
$\mu_{\sub{v}}$ is the isovector part of the nucleon magnetic moment, 
and $\mu_{\Delta{\sub{N}}}$ is the transition magnetic moment
which is proportional to the M1 amplitude (9) in the limit $k\to0$. 
This relation is trivially fulfilled in quark models, but, as shown 
already in Ref.~\cite{CGLN}, it is quite general and model 
independent. In models with only valence quarks,
the ratio of the coupling constants is $\sqrt{72/25}$ which is 
some 25\% below the value deduced from the experiment.
It is therefore not surprising that almost all models where
the valence quarks play a dominant role, predict too low
values for the photoproduction amplitudes, even if they otherwise
reproduce the nucleon magnetic moments.
Furthermore, in relativistic quark models with no pions or with a 
weak pion field, such as the CDM, the nucleon magnetic moments are 
generally underestimated; both effects may explain small
values for photoproduction amplitudes obtained in such models.
In the MIT model \cite{Donoghue} a bag radius as big as 1.4 fm was 
used to obtain the value $-102\cdot10^{-3}$~GeV$^{-1/2}$ 
for $A_{1/2}$; decreasing the radius to a more realistic 1~fm would 
bring $A_{1/2}$ down to only 1/2 of the experimental value, in 
agreement with our result for the CDM.
Possible improvements, such as the linear momentum projection, 
increase the magnetic moments \cite{Neuber} and very likely also the 
production amplitudes, though probably not to the extent that they 
would cure considerably the disagreement in this type of models.
In the LSM, the nonlinear effects from the Mexican hat
potential increase the ratio of the coupling constants to 2.05 
which agrees well with the experimentally deduced value.
Since $\mu_{\sub{v}}$ is also close to the experimental value,
this may explain good overall agreement in Fig.~3
except for the mentioned discrepancy at very low $K^2$.
This discrepancy still remains an open question; let us just mention 
that it is inherent not only to this model since a very similar 
behaviour was found in a recent calculation in the constituent 
quark model using the light-front approach \cite{Capstick}.

In summary, we conclude that the pion cloud plays an important 
role in the electroproduction of the $\Delta$ resonance.
It yields the major contribution to the quadrupole amplitudes 
which show a consistent behaviour also for $K^2\ne 0$.
Furthermore, it may considerably contribute to the absolute values 
of the production amplitudes and may, as in the LSM, yield a good 
agreement with the experimental results for $-K^2 > 0.2$~GeV$^2$.
Yet, it still does not resolve the problem why almost all theoretical 
predictions underestimate the experimental values at $K^2=0$.
We have also shown that the results at the photon point are rather 
insensitive to the details of the model; in order to be able to 
test different models it is necessary to go to nonzero $K^2$.
Here, large values of the $C2$ amplitude may be an important 
indication of the presence of a strong pion cloud in the interior 
of the nucleon and the $\Delta$.

BG and S\v{S} would like to acknowledge the hospitality they 
enjoyed during a visit to the Department of Physics of the 
University of Coimbra and MF the hospitality at J.~Stefan Institute.
This work was  supported by the Calouste Gulbenkian Foundation 
(Lisbon), the Ministry of Science of Slovenia, and the European 
Commission contract ERB-CIPA-CT-92-2287.

\end{document}